\documentclass[trackchanges,twocolumn,twocolappendix]{aastex7}

\usepackage{rotating}
\usepackage{graphicx}
\usepackage{caption}

\begin{document}

\title{Identifying Transient Hosts in LSST's Deep Drilling Fields with Galaxy Catalogues}

\author[orcid=0009-0002-9460-9900,sname='Weston']{J G Weston}
\affiliation{Astrophysics Research Centre, School of Mathematics and Physics, Queen’s University Belfast, BT7 1NN, UK}
\email[show]{jweston04@qub.ac.uk}  

\author[orcid=0000-0002-1229-2499,sname='Young']{D R Young}
\affiliation{Astrophysics Research Centre, School of Mathematics and Physics, Queen’s University Belfast, BT7 1NN, UK}
\email[]{d.r.young@qub.ac.uk} 

\author[orcid=0000-0002-8229-1731,sname='Smartt']{S J Smartt}
\affiliation{Department of Physics, University of Oxford, Keble Road, Oxford, OX1 3RH, UK}
\affiliation{Astrophysics Research Centre, School of Mathematics and Physics, Queen’s University Belfast, BT7 1NN, UK}
\email[]{S.Smartt@qub.ac.uk}  

\author[orcid=0000-0002-2555-3192,sname='Nicholl']{M Nicholl}
\affiliation{Astrophysics Research Centre, School of Mathematics and Physics, Queen’s University Belfast, BT7 1NN, UK}
\email[]{matt.nicholl@qub.ac.uk}  

\author[orcid=0000-0001-7039-9078,sname='Jarvis']{M J Jarvis}
\affiliation{Department of Physics, University of Oxford, Keble Road, Oxford, OX1 3RH, UK}
\affiliation{Department of Physics and Astronomy, University of the Western Cape, Robert Sobukwe Road, Bellville 7535, South Africa}
\email[]{matt.jarvis@physics.ox.ac.uk} 

\author[orcid=0000-0003-2265-5983,sname='Whittam']{I H Whittam}
\affiliation{Department of Physics, University of Oxford, Keble Road, Oxford, OX1 3RH, UK}
\affiliation{Department of Physics and Astronomy, University of the Western Cape, Robert Sobukwe Road, Bellville 7535, South Africa}
\email[]{imogen.whittam@physics.ox.ac.uk} 

\begin{abstract}

The upcoming Vera C. Rubin Observatory Legacy Survey of Space and Time (LSST) will enable astronomers to discover rare and distant astrophysical transients. Host-galaxy association is crucial for selecting the most scientifically interesting transients for follow-up. LSST Deep Drilling Field observations will detect distant transients occurring in galaxies below the detection limits of most all-sky catalogues. Here we investigate the use of pre-existing, field-specific catalogues for host identification in the Deep Drilling Fields (DDFs) and a ranking of their usefulness. We have compiled a database of 70 deep catalogues that overlap with the Rubin DDFs and constructed thin catalogues to be homogenised and combined for transient-host matching. A systematic ranking of their utility is discussed and applied based on the inclusion of information such as spectroscopic redshifts and morphological information. Utilising this data against a Dark Energy Survey (DES) sample of supernovae with pre-identified hosts in the XMM-LSS and ECDFS fields, we evaluate different methods for transient-host association in terms of both accuracy and processing speed. We also apply light data-cleaning techniques to identify and remove contaminants within our associations, such as diffraction spikes and blended galaxies where the correct host cannot be determined with confidence. We use a lightweight machine learning approach in the form of extreme gradient boosting to generate confidence scores in our contaminant selections and associated metrics. Finally, we discuss the computational expense of implementation within the LSST transient alert brokers, which will require efficient, fast-paced processing to handle the large stream of survey data.

\end{abstract}

\section{Introduction}
The Vera C. Rubin Observatory's Legacy Survey of Space and Time (LSST) Deep Drilling Field (DDF) mini-survey will observe five specific Rubin pointings with an enhanced cadence. These pointings have been selected due to the extensive amount of pre-existing data in the regions from prior surveys carried out over the past thirty years. With a higher cadence and greater depth than the wide-fast-deep survey, the DDF program will generate a greater number of transient alerts per square degree each night, making it a well-suited avenue for high-redshift transient science and cosmology \citep{Abbott2024, dmtn-102, Gris2024, Hambleton2012, Scolnic2018}. The rate of transient alert output (an estimated 10 million per night) will require extensive filtering, both to separate real candidates from fake detections and to distinguish more `interesting' transients from common ones \citep{ldm-612}. 

Efficient identification of real extragalactic transients and the subsequent selection of those meriting follow-up observations requires methods of `host-matching' wherein we determine the galactic host within which the transient occurred. With this additional contextual information, we can infer the reality of a transient, its redshift, and constraints on its likely spectroscopic or physical type. In the DDFs, we expect to perform host-matching thousands of times each night.

Multiple catalogue-based methods of transient host-matching exist. A simple angular approach calculates the angular separation between a transient and its potential hosts to determine the `best' host as the one with the shortest distance \citep{Qin2022}. A redshift-based approach can be used to estimate physical distance from a given transient, provided redshift measurements are high-quality \citep{Qin2022}. A morphology-based approach utilises structure parameters such as galaxy axis lengths and position angle to determine a transient's separation from the host in units of the galactic radius \citep{Sullivan2006,Gupta2016}. Image-based approaches utilise processed survey images in models such as Convolutional Neural Networks (CNN) to identify key associated structures \citep{Gagliano2021,Forster2022}. Implementations of transient host-matching algorithms typically use a combination of the above approaches, depending on the amount of contextual data available near the transient location. In current alert brokers, such as Lasair \citep{Williams2024}, the cross-matching algorithm will incorporate wide-sky pre-existing catalogue data from prior surveys. Pre-existing wide-field catalogues such as Pan-STARRS and the Sloan Digital Sky Survey (SDSS) are well matched to the depths reached by all-sky transient surveys and can therefore be used to identify the correct host for the majority of transients.

The current implementation of these pipelines may encounter issues with the depth of the LSST Deep Drilling Fields. It is unlikely that existing wide-field catalogues will contain the fainter and more distant galaxies that will be seen with Rubin. Because the DDFs reach several magnitudes deeper than the Wide Fast Deep (WFD) survey, the surface density of detectable galaxies increases by more than an order of magnitude. As a result, any transient in a DDF will have far more potential hosts within a small angular radius than transients discovered in the WFD survey. This greatly increases the risk of ambiguous or incorrect associations unless deeper, field-specific catalogues are used. Until the first LSST data release, which is expected approximately 1.5 years into the survey, we will have no wide-field catalogues that match the depth of the DDFs. However, the Deep Drilling Fields were selected by LSST for their pre-existing deep coverage over a small area. In selecting the useful information from these existing narrow-field catalogues, we may be able to perform host-matching at LSST depths within the DDFs from the beginning of the survey. 

Seven independent Rubin alert brokers are currently designed to filter the LSST transient streams for a variety of science goals and provide contextual information, removing bogus transient candidates and promoting those most interesting for further science \citep{ldm-612,Matheson2021,Moller2021,Forster2021,Williams2024}. Most involve some cross-matching with internal or external archival data. Real-time transient processing for large-scale surveys is a computationally expensive task, with LSST expected to detect thousands of extragalactic transient events each night in the WFD survey. Any changes to this cross-matching - whether to refine the methods of host association in the deep drilling fields or to incorporate additional deep drilling field data - must consider the impact on the broker pipeline's efficiency. 

This paper outlines the selection of pre-existing catalogues within the Rubin DDFs for their use in transient-host matching and contextual classification in Rubin-LSST. We present a repository of the catalogues and the requisite Python code to produce summary analytics for ranking each set of data, before selecting the best catalogues for testing with samples of transients from the Dark Energy Survey (DES) in three of the Deep Drilling Fields. We compare a method that determines transient-host separation in units of the host's semimajor axis against the computationally more expensive Directional Light Radius (DLR) method \citep{Sullivan2006,Gupta2016}. We discuss the performance and cost of implementation within transient alert pipelines for LSST.  Finally, we develop machine-learning models to generate confidence scores for our associations and key features to indicate the reliability of a host match.

Following an introduction to LSST in Section 2, we give an outline of the transient alert brokers and contemporary methods for host matching in Section 3. Section 4 outlines the use and scope of the Deep Drilling Field catalogues, as well as our selection and ranking of the datasets. In Section 5, we apply host-matching methods to these datasets using samples of transient data from the Dark Energy Survey (DES), Hyper Suprime-Cam (HSC) and the Asteroid Terrestrial Impact Last Alert System (ATLAS). Finally in Section 6 we discuss the implementation of these catalogues and the attempted association techniques within the alert brokers. 

\section{LSST \& The Deep Drilling Fields}
The Vera C. Rubin Observatory's Legacy Survey of Space and Time (LSST) is a ten-year all-sky survey aiming to cover the 19,600 square degree southern sky approximately 800 times with rapid revisit timescale requirements \citep{LSST2009,Ivezic2019,PSTN-056}. The current observing strategy is set by the Survey Cadence Optimisation Committee (SCOC) Phase 2 recommendations. The 3.2 Gigapixel camera obtains observations in the six \textit{ugrizy} filters, providing coverage between 320 and 1035nm. A single LSST visit will reach a limiting magnitude $\sim$24-25 in the g, r, or i bands (or somewhat shallower in the other bands); for co-added images this depth approaches r$\sim$28. It is expected that 30 Terabytes of nightly data will be generated.

The Rubin survey strategy divides its time between several types of observations. The ``Wide Fast Deep" (WFD) survey utilises 80\% of the total survey time to observe the majority (19,600 square degrees) of the southern sky in areas with low-dust extinction or high stellar density. Mini-surveys target the North Ecliptic Spur, the dustier portions of the Galactic Plane and the Southern Celestial Pole. A smaller Near-sun Twilight Microsurvey focuses observations around the ecliptic plane during Twilight. Recommendations suggest that up to 3\% of Rubin's survey time be available for Target of Opportunity observations; the details of implementation are still under discussion.
\subsection{The Deep Drilling Fields}
The Deep Drilling Field (DDF) program consists of five Rubin pointings observed at a high cadence with 30-40 visits per night over a total area of 57.6 square degrees \citep{Scolnic2018,Gris2024}. The program provides several advantages for Rubin transient science. For a broader view of transients, the higher cadence will produce excellent light curves in six bands (all six being observed each night) that will help us better understand the physics of low-to-high redshift supernovae, nuclear transients, and especially fast transients. The Dark Energy Science Collaboration aims to leverage the Type Ia supernovae data to extend the SN Hubble diagram by a factor of 2 past the WFD. Observations of Type Ia Supernovae in the DDFs will provide calibration for the high-quality light curve templates required to fit the Rubin WFD data. The DDFs are chosen both for their large amounts of archival data from previous surveys across the spectrum and for their absence of bright foreground sources that cause ghosting, detector saturation, bleeds, and diffraction spikes. They consist of:

\begin{itemize}

    \item COSMOS; the Cosmic Evolution Survey Field \citep{Scoville2007}. The Cosmic Evolution Survey includes X-Ray to Radio imaging from a range of space (\textit{Hubble}, \textit{Spitzer}, \textit{XMM}, \textit{Chandra}, among others) and ground-based telescopes (Keck, Subaru, MeerKAT, VISTA, etc) \citep{Finoguenov2007,Sanders2007,Schinnerer2007,Taniguchi2007,McCracken2012,Jarvis2013,Civano2016,Heywood2022,Weaver2022}. It is the most extensively covered of the DDFs and continues to be observed by these other instruments.

    \item ELAIS-S1; European Large-Area ISO Survey-S1 \citep{Oliver2000}. ELAIS-S1 was the largest targeted survey by ESA's Infrared Space Observatory using the ISOCAM and ISOPHOT instruments. Extensive archival data exist at radio, optical, NIR, and X-ray wavelengths \citep{LaFranca2004,Berta2006,Puccetti2006}.

    \item XMM-LSS; the XMM-Large scale Structure \citep{Pierre2004}. The XMM-LSS field has been observed by XMM-Newton in X-ray wavelengths. Subsequent follow-up observations of the field yield coverage up to radio, including CFHT (Canada France Hawaii Telescope), Spitzer, the Low-Frequency Array (LOFAR) and Subaru \citep{LeFevre2004}. 

    \item ECDFS; the Extended Chandra Deep Field-South or W-CDF-S (Wide Chandra Deep Field-South)\citep{Lehmer2005,Xue2016}. The Chandra Deep Field-South, originally observed by the Chandra X-Ray Observatory, subsequently underwent extended-field follow-up observations by Chandra, HST, VLT, and Spitzer, among others \citep{Cardamone2010,Silverman2010}. 

    \item EDF-S; the Euclid Deep Field-South \citep{Scarlata2019,PSTN-056}. EDF-S has less multi-wavelength data compared to the four DDFs above, and so was finalised as a collaborative observation between Rubin and the Euclid survey. The aim is to combine LSST observations with Euclid's high-resolution optical/NIR imaging to yield better-calibrated photometry and more accurate photometric redshifts for cosmology \cite{Rhodes_2017,Euclid2022}.

\end{itemize}

The Phase 3 recommendations of the LSST SCOC are that 6-7\% of the total survey time be dedicated to the DDFs, with each receiving approximately 20,000 nightly visits over 10 years \citep{PSTN-056, Scolnic2018}. The exception to this is the COSMOS field, which will receive 20,000 visits by the end of Y3. There will then be a `complete' DDF with which to address the program's science goals. EDF-S consists of two Rubin pointings but will be observed over its entire area in 20,000 visits, and as such, will only achieve half the depth of the other DDFs.

\begin{figure}
\centering
 \includegraphics[width=1\columnwidth]{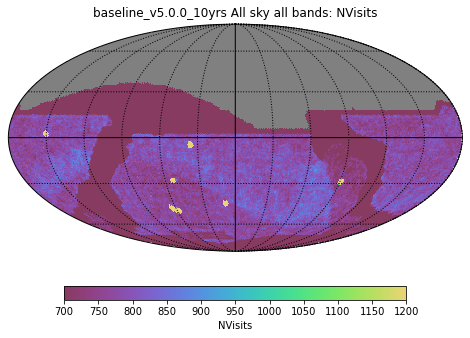}
 \caption{Pointings for the DDF mini-survey. Although the scale saturates at 1200 visits, the DDF pointings, clearly visible in yellow, will each receive 10,000-20,000 visits \citep{Yoachim2025}.}
 \label{fig:0201}
\end{figure}

The DDF pointings are summarised in Table \ref{tab:211}. For EDF-S, observations are divided between the two pointings EDFS-a and EDFS-b. Enlarged pointings are displayed for reference in Figure \ref{fig:0201}.

\begin{table*}[ht]
\centering
\caption{Deep Drilling Field Locations}
\label{tab:211}
\begin{tabular}{lcccccc}
\hline
Field   & ELAIS-S1 & XMM-LSS & ECDFS  & COSMOS & EDFS-a & EDFS-b \\
\hline
RA [deg]       & 9.45    & 35.57    & 52.98   & 150.11  & 58.9    & 63.6    \\
Dec [deg]      & -44.02  & -4.82    & -28.12  & 2.23    & -49.32  & -47.6   \\
Gal $l$ [deg]  & 311.29  & 171.1    & 224.07  & 236.78  & 257.9   & 254.48  \\
Gal $b$ [deg]  & -72.88  & -58.91   & -54.6   & 42.13   & -48.46  & -45.77  \\
Eclip $l$ [deg] & 346.66 & 31.59    & 40.81   & 151.39  & 32.0    & 40.97   \\
Eclip $b$ [deg] & -43.2  & -17.92   & -45.44  & -9.34   & -66.61  & -66.6   \\
\hline
\end{tabular}
\end{table*}

The program provides several advantages for Rubin transient science. For a broader view of transients the high-cadence will produce excellent light curves in six bands for understanding the physics of low-to-high redshift supernovae, nuclear transients, and especially fast transients. The Dark Energy Science Collaboration aims to leverage this Type Ia supernovae data to extend the SN Hubble diagram by a factor of 2 past the WFD. Observations of Type Ia Supernovae in the DDFs will provide calibration for the high-quality light curve templates required to fit the Rubin WFD data.
\section{Transient Host Matching}
Cross-matching refers to the process of determining nearby `associated' records in pre-existing data using observational data for a given object, such that historical information can inform our present understanding of the object, and vice versa. For example, a detected transient's coordinates may be matched against a known CV (cataclysmic variable) or an AGN (active galactic nucleus). Host matching refers to cross-matching of an extragalactic transient (such as a supernova) with an associated host galaxy. Contextual classification, in this case, refers to using the cross-matched data for an object to infer its reality and constrain its type. The detail of this classification can range from a simple real/bogus separation, to distinguishing stellar from nuclear transients, and at its most precise, to differentiating between specific supernova types.

Different supernova types are defined by their spectra, which can only be obtained through follow-up observations. While photometric analysis of the supernova lightcurve provides estimates of an object's redshift and type, these estimates are much less reliable, and the inclusion of photometrically classified supernovae can contaminate spectroscopic sample data. In both cases, additional contextual information can assist in the ultimate classification; the secure identification of a host with a spectroscopic redshift further assists in the classification of the transient. Global host galaxy properties such as stellar mass, age, metallicity and star-formation rates can provide context for a transient's evolution and (in the case of supernovae) explosion mechanisms \citep{Anderson2010,Wiseman_2020}.

As LSST returns a high number of potential transients each night, the allocation of follow-up observations will become more competitive. The number of LSST transients that can be followed up spectroscopically will be orders of magnitude smaller than the number of transients discovered. It is therefore not only important that all available data is utilised, but that the method of cross-matching strikes the right balance between efficiency and accuracy to appropriately process the data. 

Several approaches to cross-matching currently exist. The simplest is a spatial cross-match, which returns the source with the smallest angular separation from the transient location. The object with the lowest angular separation is then selected as the most likely cross-match. This is reasonable when the angular separation is so low that the transient \textit{is} confidently identified as being spatially coincident with the catalogued object, as with AGN or variable stars.

For extragalactic transients outside of galaxy centres, a morphology-based approach may be better motivated. If we have information on the size and shape of a cross-matched galaxy (e.g. axis lengths and position angle in the direction of the transient), we can evaluate the likelihood of association based on the projected distance along the galaxy's morphology \citep{Sullivan2006,Gupta2016}. If redshift information is available, we can also convert the angular separation to a physical distance and select only galaxies with a physically plausible distance for association. For truly accurate physical distance measurements, we must use either high-accuracy photometric redshifts or spectroscopic redshifts. 

In an ideal scenario, we are able to combine these approaches, dependent on need, to reach a given accuracy threshold for our host selection. A host identified via a physical approach with a low-accuracy redshift may be verified via a morphology-based approach. In low-density areas, it may even be best to utilise a spatial approach as a lighterweight method.
\subsection{Sherlock}
Sherlock \citep{Young2023} is a Python package and MariaDB database of catalogues utilised  by the Lasair transient alert broker \citep{Williams2024}, the ATLAS and Pan-STARRS survey pipelines \citep{Smith_2020,Stevance2025,Fulton_2025} and the Public ESO Spectroscopic Survey for Transient (PESSTO) Marshall \citep{Smartt_2015}. Using a set of all-sky catalogues it provides spatial cross-matching for a given astrophysical object to aid in contextual classification.

Sherlock uses the catalogues to identify transients that are synonymous with a catalogued source. These include variable stars (VS), known cataclysmic variables (CV), known Active Galactic Nuclei (AGN) and other transients found in the centre of galaxies (Nuclear Transients, NT). These transients fall within the ``synonym radius'' of a catalogued source of the same type. Sherlock also attempts to associate non-synonymous transients with their catalogued counterparts. A detection falling within an ``association radius'' of a catalogued star above a certain brightness will be classified as an artefact related to a bright star (BS). In contrast, a detection found in association with a catalogued galaxy will be classified as a supernova (SN). If Sherlock fails to match the transient against any catalogued source, it is classified as an Orphan.

The synonym radius used in classifications of VS, CV, AGN, and NT is mainly determined by the resolution of a given survey's images. The association radius is more dependent on metrics such as the brightness of a catalogued source or (in the case of galaxies) its angular size. For potential hosts with morphological data, associations are rejected if a transient is separated from a galaxy by more than 2.4 times the galaxy's semimajor axis radius. In the case where there is a distance measurement or redshift for a galaxy, allowing us to calculate physical separations between the host-transient pair, Sherlock rejects matches where the distance is greater than 50 kiloparsecs (Srivastav et.al., in prep).

Sherlock performs contextual classification of sources to return a list of possible synonyms/associations for each transient. A ranking algorithm is then used to select the most likely classification.

Sherlock's database currently contains datasets from a number of all-sky surveys and source-specific catalogues. All-sky surveys include data from DESI Legacy Imaging Survey DR10 \citep{Dey2019}, Gaia DR3 \citep{Gaia2018}, Pan-STARRS \citep{chambers2019,Tachibana2018}, SDSS DR12 \citep{Alam2015}, 2MASS \citep{Skrutskie2006} and the Guide Star Catalogue \citep{Lasker2008}. For the source-specific data, Sherlock utilises catalogues such as the Million Quasars Catalogue \citep{Flesch2023}, the Downes CV Catalogue \citep{Downes2001}, and the Ritter Cataclysmic Binaries Catalogue \citep{Ritter2003}. In addition to the source catalogues above, Sherlock also contains spectroscopic and distance measurements from the NED-D Galaxy Catalogue \citep{Steer2017}, the NASA/IPAC Extragalactic Database via \cite{Young2023b} and the LASr-GPS volume-limited galaxy catalogue \citep{Asmus2020}. More catalogues are added to the library as they become publicly available; currently, the database consumes approximately 4.5TB.

Currently, the above catalogues are well matched to wide-field, relatively shallow surveys. In the DDFs, we require adjustments - firstly, to account for the greater depth, field-specific catalogues must be implemented for associations to be achieved for higher-redshift sources. A large amount of data exists for the DDFs to achieve this. Secondly, to account for the more crowded nature of the catalogue data in the DDFs, the synonym and association radii, as well as cross-matching methods, will need to be reviewed to ensure the software maintains a satisfactory level of accuracy in the new environment. 
\subsection{Host Matching in LSST}
Rubin Alert packets are created via the detection of all sources in the corresponding LSST difference image with a \texttt{SNR} greater than 5; it is predicted that 10 million total alerts per night will be produced. Not all of these will be real objects - with LSST expecting to discover 2 million quasars, 10 million supernovae, and 50 million variable stars across the ten-year survey, the ratio of alerts to real objects may be high \citep{dmtn-102}.

The Rubin alert brokers are a series of seven independent projects designed to filter the LSST transient streams for different science goals and provide additional contextual information \citep{ldm-612}. Each broker has been developed independently, but typical broker tasks might include filtering alerts based on their content and history, and performing cross-matching with archival data. This filtered/enhanced data stream can then be used to identify and prioritise objects for follow-up observation, with or without a classification step.

The Lasair science platform \citep{Williams2024} is the UK Community Broker for transient alerts. For host-matching and contextual classification, Lasair uses Sherlock to cross-match against the pre-existing data as detailed in the previous section.


\subsection{Morphology-based association}
A morphology-based association determines the most likely host using galaxy size, shape and position angle on the sky (measured east through north). By calculating which galaxy is `most' extended towards the transient, we can identify it as the most likely host association. 

Calculation of the axis lengths and position angles of galaxies is typically done through software such as Source-Extractor \citep{Bertin1996}. The program builds a catalogue of astronomical sources from isophotal measurements of survey images. In calculating these measurements for an elliptical shape, Source-Extractor calculates the semimajor and semiminor axes \texttt{A} and \texttt{B} with the position angle \texttt{THETA}. \texttt{A} and \texttt{B} are the maximum and minimum spatial dispersion of the object profile along any direction. This dispersion and the `edge' of a galaxy are dependent on the detection threshold, often expressed in units of root mean square (RMS) noise. A lower threshold accounts for fainter regions and can lead to larger radii.

Morphological crossmatching approaches can incorporate a range of details for higher accuracy at the cost of computational speed. The standard and straightforward method is to treat the elliptical shape of the galaxy as being strictly circular; i.e., the galaxy is equally extended in all directions (either at radius \texttt{A} (semi-major axis length) or \texttt{B} (semi-minor axis length), an average of the two, or a mean half-light radius). We calculate the angular separation between the transient and the galactic centre and express it in units of this effective radius to give a dimensionless distance. In subsequent sections of this paper, we calculate the distance using the semimajor axis \texttt{A} and denote this calculated distance as the `A-Value'. 

A more sophisticated approach is to use the Directional Light Radius (DLR) method \citep{Sullivan2006,Gupta2016}. Here we calculate the directional radius of the galaxy along the line connecting the transient position to the host centre. Then the Directional Light Radius along this direction is given as:

\[
\phi = \arctan2(\Delta y, \Delta x) - \theta
\]

\[
\mathrm{DLR} = \frac{a \, b}{\sqrt{(a \sin\phi)^2 + (b \cos\phi)^2}}
\]

where $\theta$ is the position angle measured east through north, and \(\phi\) is the angle from the galaxy's major axis to the line connecting the galaxy centre to the transient. We can calculate the dimensionless ratio $d_{\mathrm{DLR}}$ as

\[
d_{\rm DLR} = \frac{\text{Transient--Host angular separation (arcsec)}}{\mathrm{DLR}}
\]

In both approaches - DLR and A-Value - we identify and select the lowest value as our `best candidate' host.

\begin{figure}
\centering
 \includegraphics[width=0.8\columnwidth]{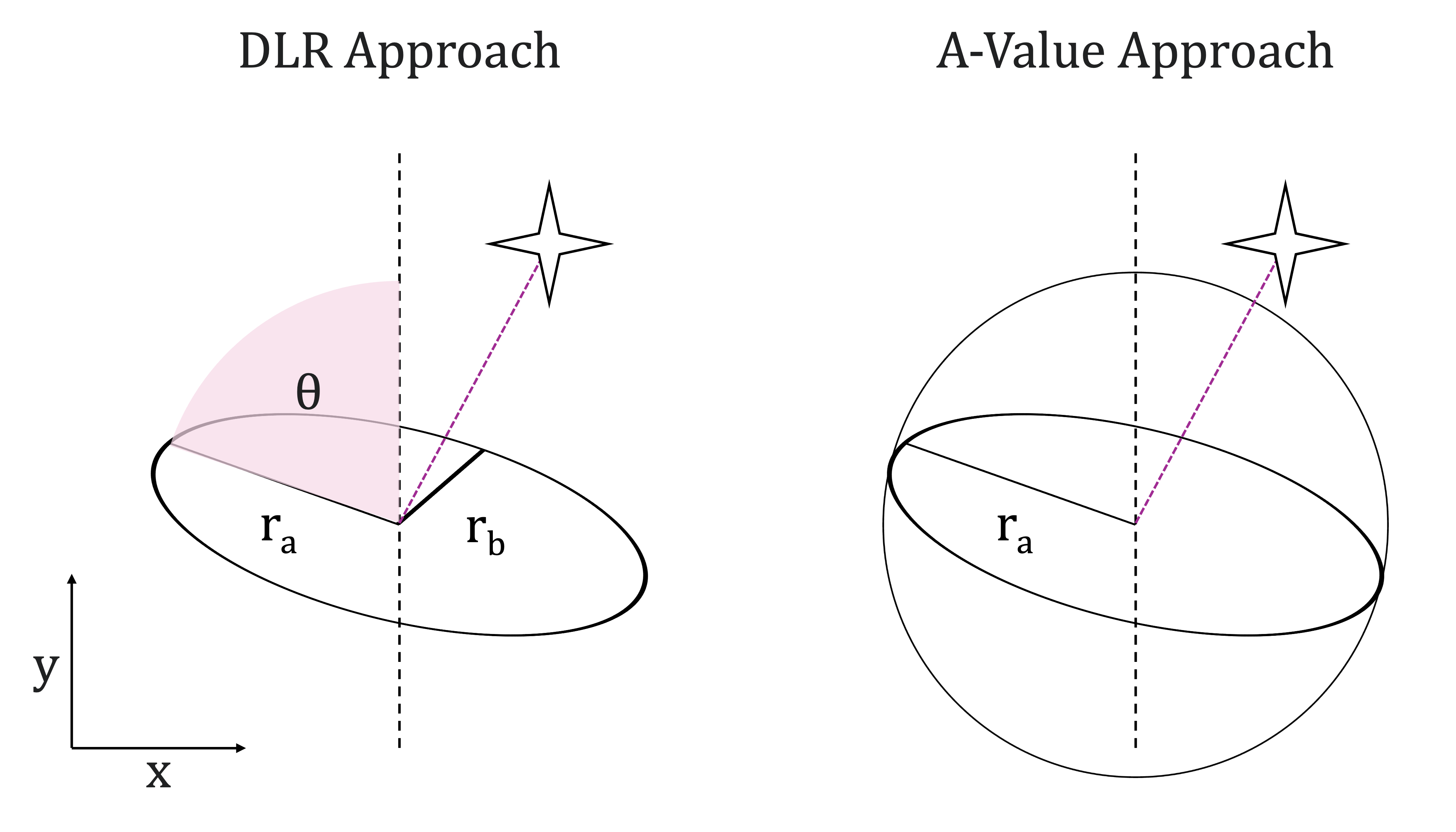}
 \caption{Comparison of the DLR and A-Value methods. In the A-Value method, a less detailed morphology creates `larger' galaxies from which to perform cross-matching.}
 \label{fig:0331}
\end{figure}

As we have seen in the case of Sherlock, these approaches are not strictly limited to identifying the host galaxy of a transient. While objects such as stars typically lack morphological parameters, the position of a transient relative to its galactic host can inform our understanding of its type. For example, a low DLR can be used to infer the transient is nuclear; an AGN, or Tidal Disruption Event (TDE). Outside the galactic nucleus, we can determine the likelihood of a supernova occurring at a given distance from its selected host centre; outside of a given $d_{\mathrm{DLR}}$, we can infer that the transient is hostless.

The DLR method requires significantly more steps in its calculation than the A-Value method. As such, when discussing implementation, it is important to consider the benefits to host-matching accuracy gained from this approach. For instance, an immediate benefit is that the DLR method is less biased towards large, high-ellipticity galaxies, which the A-Value method would treat as circular. 

It is also important to consider -- particularly when implementing this approach across multiple catalogues -- that morphological approaches using isophotally-derived metrics (i.e., radii and position angles) produce survey-dependent quantities. Images of galaxies and the subsequently derived axes and position angles depend on the observing filters, PSFs, survey depth or sensitivity, and the assigned detection threshold or chosen isophote. While we see little impact of these differences in our testing, it remains essential to consider and record the catalogue from which the best associations for a given transient were obtained.

\cite{Gupta2016} takes quantities used in calculating $d_{\mathrm{DLR}}$ (e.g. ratio between the lowest values, minimum magnitude, ellipticity, galaxy magnitude) and trains a machine learning model to generate confidence scores for given associations. It is important that an association not only be the best available, but also reliable. While generating these scores is expensive in a live pipeline, offline machine learning analysis would enable assessment of association accuracy. We would expect that, in crowded fields with multiple nearby galaxies, the association would be assigned a lower confidence score. In the case of blended hosts, for example, it is difficult to determine without follow-up observation which galaxy the transient belongs to, even if the calculated $d_{\mathrm{DLR}}$ is low.
\section{The Deep Drilling Field Catalogues}
We have previously discussed the extensive pre-existing coverage of the DDFs. These catalogues typically summarise small, field-specific studies that go deeper than all-sky surveys. These catalogues allow us to examine fainter, higher-redshift galaxies and other objects, and often provide multi-wavelength data. In this section, we collect available data from deep catalogues overlapping the LSST DDFs and assess the utility of each catalogue for cross-matching in LSST alert brokers.

Catalogues were obtained from a number of sources, including VizieR via the Strasbourg astronomical Data Centre (CDS) service \citep{Genova2013}, SIMBAD \citep{Wenger2000}, CosmoHub \citep{Carretero2017, Tallada2020}, and The Spitzer Spectroscopic Data Fusion \citep{Vaccari2025}. As of writing, we have collected a total of 67 catalogues from various surveys across all fields, except Euclid-South. Of these, 56 contain more than 1,000 entries and 44 contain more than 10,000 entries. Of these 44, 28 contain redshift measurements (spectroscopic or photometric). For the morphology-based approach, we identify 15 catalogues across COSMOS, ECDFS, and XMM-LSS that provide detailed spatial extent information (Semimajor axis length, semiminor axis length, and position angle). More catalogues contain other morphology data, such as the effective radius. Finally, for contextual classification, we identify 29 catalogues with classification flags for different astronomical objects, such as Star/Galaxy, AGN, and QSO flags. A complete summary of the catalogue collection and coverage plots can be found in Figures \ref{fig:0401} and \ref{fig:0403}, and in Tables 2, 3, 4, and 5 in the appendices.

We have neglected to include a table for EDFS, which so far is covered by the Euclid Quick Data Release 1 \citep{euclidqr1}. This release contains a photometric redshift and spectroscopic catalogue for the three Euclid deep fields, including EDF-South.

\begin{figure}
\centering
 \includegraphics[width=1\columnwidth]{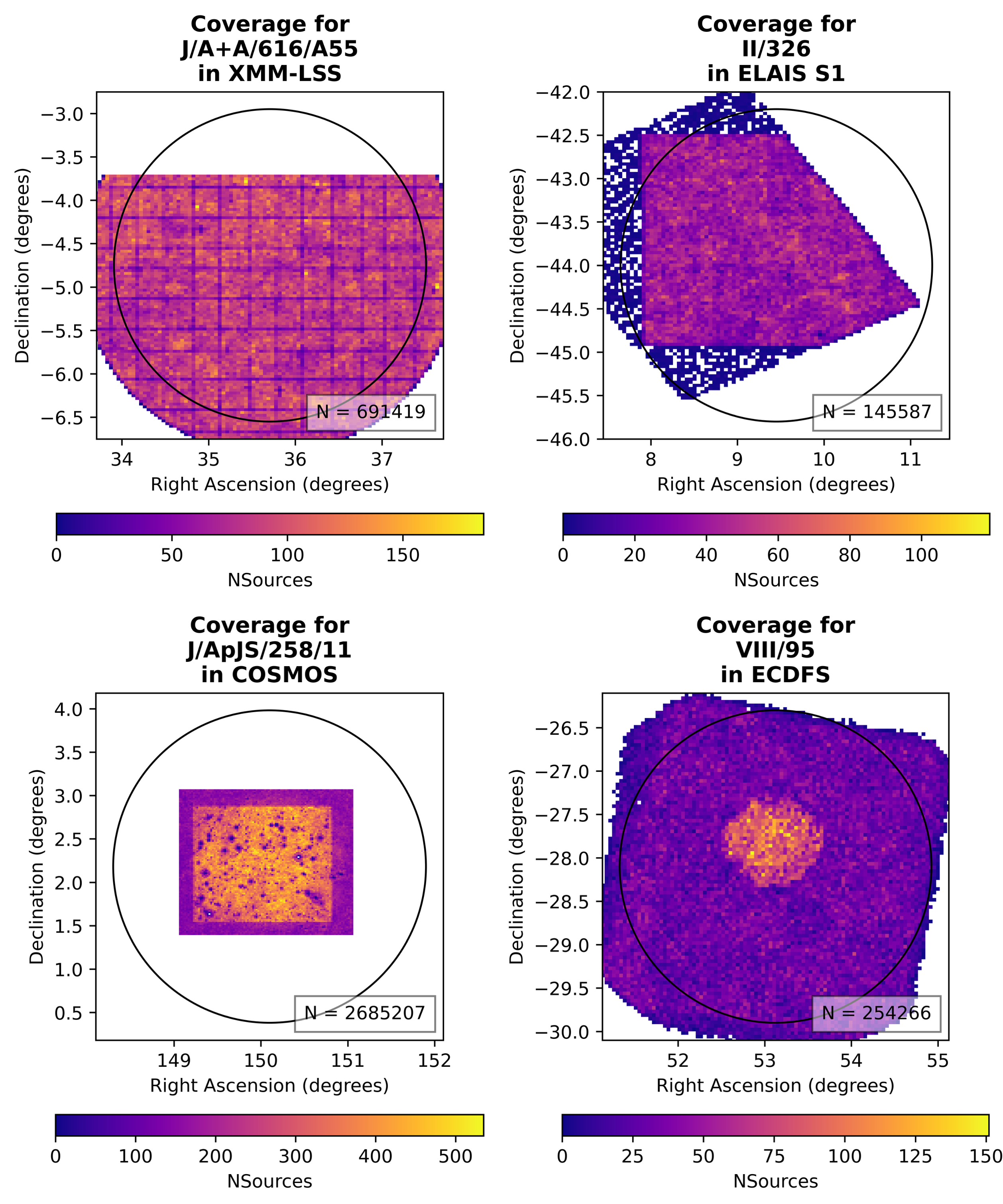}
 \caption{Example field coverage plots for a DDF catalogue in four of the fields. Lestrade performs a conesearch in the region surrounding the central pointing for each catalogue to identify subsets of data for each field. Top left: the CFHQSIR survey \citep{Pipien2018} in XMM-LSS. Top right: Revised SWIRE Photometric Redshift Catalogue in ELAIS S1 \citep{Rowan-Robison2013}. Bottom left: The COSMOS2020 Catalogue in COSMOS \citep{Weaver2022}. Bottom right: The Herschel Multi-tiered Extragalactic Survey catalogue \citep{Hermes2017} in ECDFS.}
 \label{fig:0401}
\end{figure}

\begin{figure}
\centering
 \includegraphics[width=1\columnwidth]{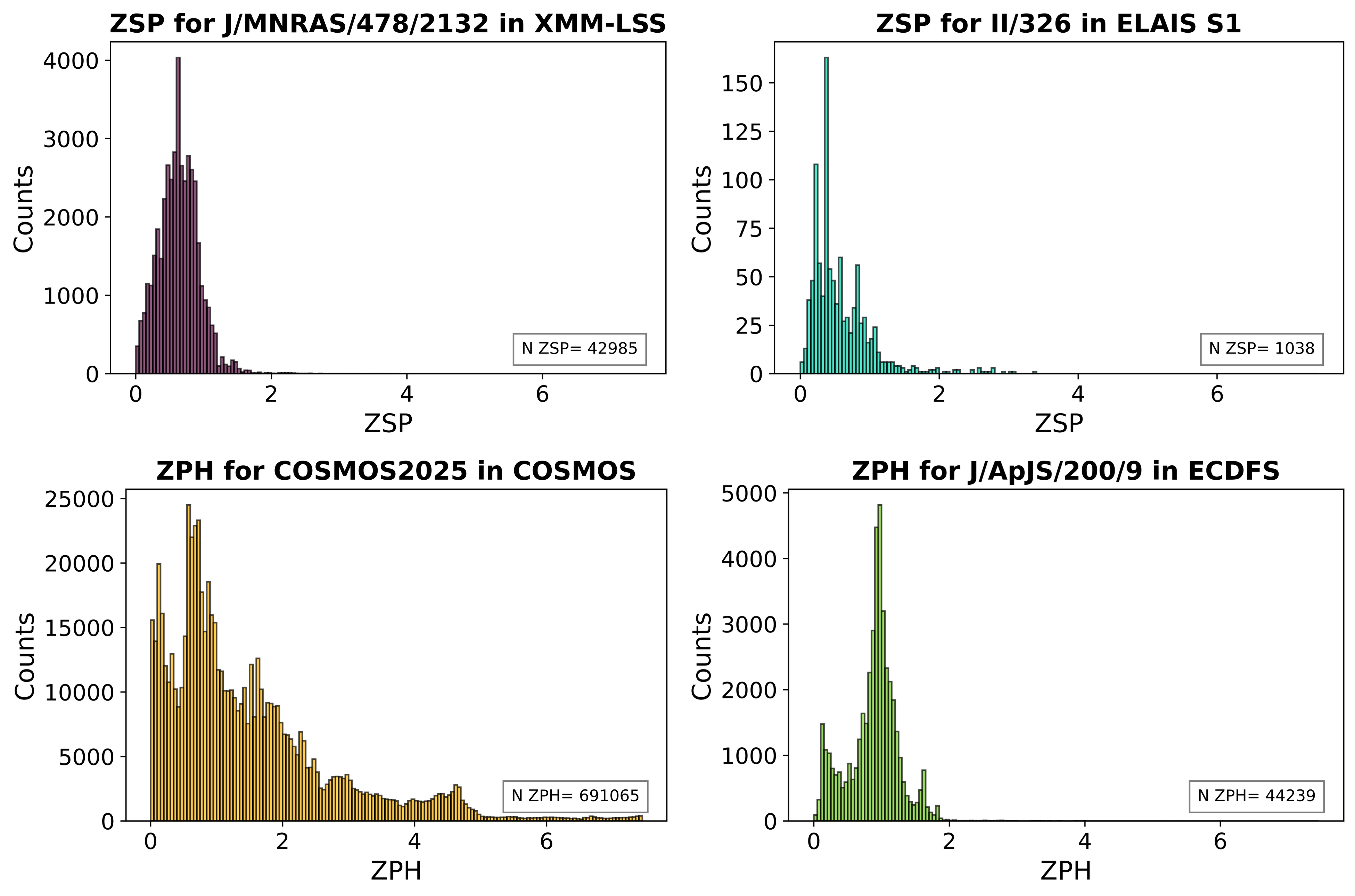}
 \caption{Example redshift coverage plots for a DDF catalogue in four of the fields. Top left: Spectroscopic redshifts in the XMM-SERVS survey \citep{Chen2018} in XMM-LSS. Top right: Spectroscopic redshifts in the Revised SWIRE Photometric Redshift Catalogue in ELAIS S1 \citep{Rowan-Robison2013}. Bottom left: Photometric redshifts in the COSMOS2025 Catalogue in COSMOS \citep{shuntov2025}. Bottom right: Photometric redshifts in the ACS-GC catalogue \citep{Griffith2012} in ECDFS.}
 \label{fig:0403}
\end{figure}

We have developed a software package, Lestrade, to process these catalogues. Lestrade provides two options for catalogue analysis. CDS bibcode-IDs may be input for automatic analysis of the corresponding README file for authors, summaries and schema. The software generates Markdown files, which are published to the repository wiki. From the schema, the software determines key spectroscopic and photometric redshifts, the existence of morphological or class-related metrics, and coverage within each DDF. These catalogue summaries can then be adjusted and re-entered by users back into Lestrade. Alternatively, the key metrics and information can be input manually by the user in the first instance to generate a more detailed Wiki page with plots for the relevant redshift metrics, redshift quality metrics and type flags.

Within this collection of catalogues, we can sort and rank each dataset by the quality and depth of its information. In ranking them, we prioritise catalogues for implementation within the cross-matching Sherlock service of Lasair.

\begin{itemize}
    \item\textbf{A+}: Spectroscopic redshift catalogues with additional contextual information. Spectroscopic redshifts provide the most reliable estimation of host-transient separation and are a priority for implementation within cross-matching pipelines. There is high value in the contextual information regarding redshift and the type of galaxy (or any other object) the transient has been cross-matched with. Redshift and luminosity distance are extremely valuable for the rapid calculation of absolute magnitudes of transients.

    \item\textbf{A}: Spectroscopic redshift catalogues with no additional contextual information. While not as well-suited to our morphology-based approach, the spectroscopic redshifts still provide significant value for cross-matching studies.

    \item\textbf{B+}: Photometric redshift catalogues with additional contextual information. While not as reliable for cross-matching, photometric redshifts still provide useful context, both individually and when used in tandem with non-luminosity-distance approaches. 

    \item\textbf{B}: Photometric redshift catalogues with no additional contextual information. In these cases, a cross-match using photometric redshifts may be less reliable and would be improved by utilising the additional contextual data from another catalogue.

    \item\textbf{C+/C}: Photometric catalogues. Such catalogues are less valuable than those containing redshift information and are therefore the lowest priority for implementation.
\end{itemize}

This ranking is not exhaustive and exists only to provide a broad overview of considerations regarding catalogue implementation, which increase in complexity when discussing the implementation of multiple catalogues. A significant omission is that of catalogue size, depth and spatial coverage.

The full ranking of the catalogues above can be found on the Lestrade wiki \footnote{https://github.com/joshgithubbin/Sherlock-DDF/wiki}, with each dataset separated by its usefulness in crossmatching. New catalogues are continually being added to the repository as they are published. 

\section{Cross-Matching in the Deep Drilling Fields}
\subsection{Cross-Matching in XMM-LSS \& ECDFS}
We test our catalogue cross-matching in the DDFs using a sample of supernovae from the Dark Energy Survey (DES) observed over a 5-year period within the XMM-LSS, ECDFS and ELAIS-S1 deep drilling fields \citep{Abbott2016,Abbott2024}. The sample contains accurate coordinates for each transient, along with manually confirmed hosts identified using the DLR method. Spectroscopic redshifts for these host galaxies were then obtained primarily from OzDES, along with other sources \citep{Yuan2015,Childress2017,Lidman2020}. The sample contains 8274 transients divided between the three DDFs, with a redshift range of $0.0001 < z < 6.84$ and a median redshift of 0.56. The DES Supernova program was carried out to a limiting magnitude of $\sim24$ in all bands; the expected median $5\sigma$ point-source depth for LSST Y1 will be $\sim26$. We process each DES transient through our own host-matching pipeline and compare the identified galaxies to the host galaxy provided in the DES sample to provide a performance measure for both the DLR and A-Value methods.

Lestrade constructs ``primary catalogues" that combine data within the specified region from the available published datasets. For morphological host-matching, we only require the host coordinates, axis lengths, and position angle. Thus, six catalogues with a wide range of data spanning multiple fields can be combined and condensed into a single five-column catalogue covering a single deep drilling field. The pipeline takes a list of the catalogues to be imported alongside any `calibrations' required to make the data homogeneous. If the morphological data are provided in units of pixels, for instance, an instrument axis scale can be provided to allow for the necessary conversions between units. 

\begin{figure}
\centering
 \includegraphics[width=0.8\columnwidth]{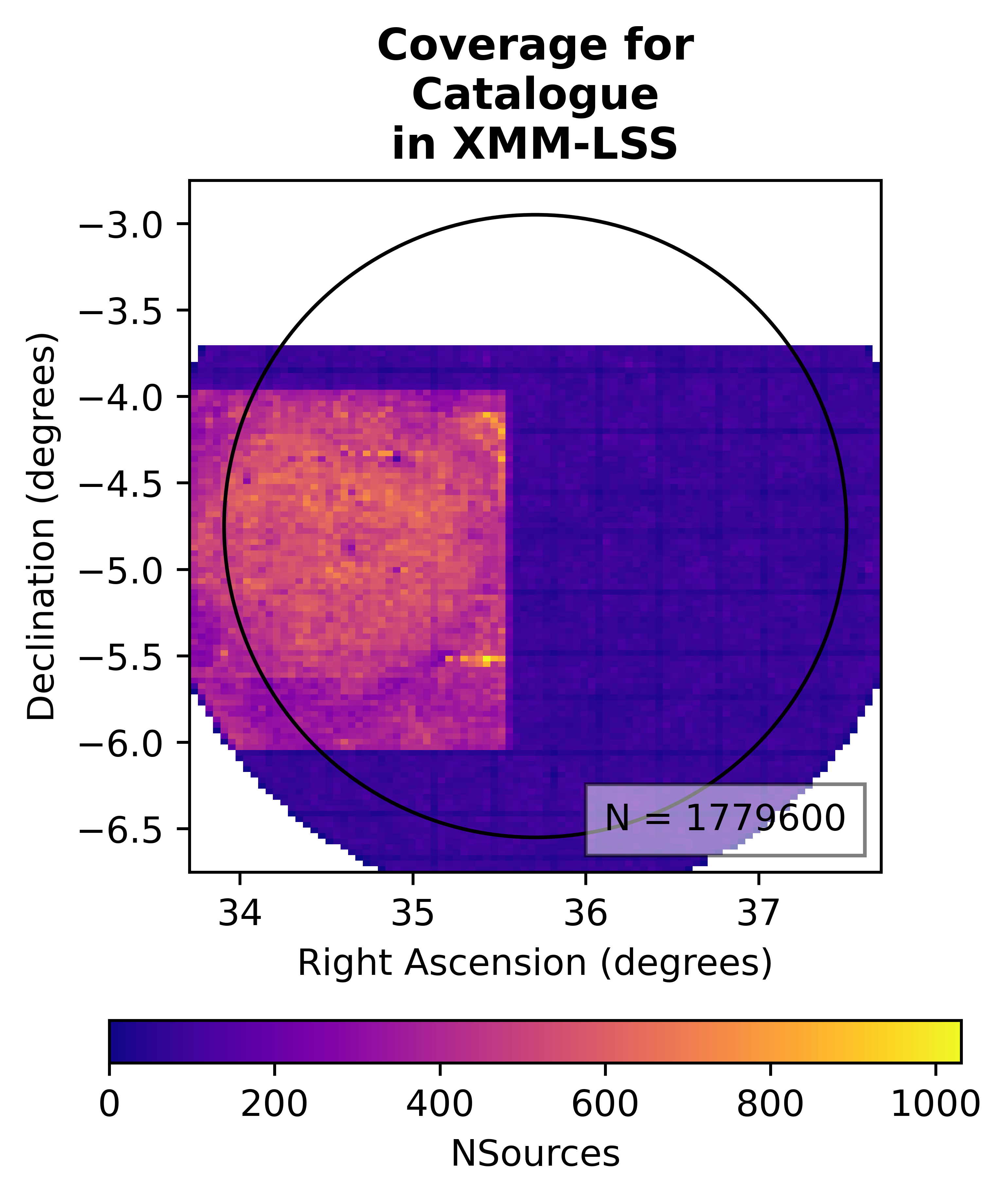}
 \caption{Primary catalogue coverage for XMM-LSS.}
 \label{fig:0510}
\end{figure}

Of the three DDFs covered by the DES sample, we lacked ELAIS-S1 catalogues with morphological information at the time of testing. We identify XMM-LSS as the best case study, with extensive pre-existing catalogue coverage and 2664 transients to test cross-matching approaches. From our catalogue coverage, we select two datasets within the region as suitable catalogues for testing. The SPLASH-SXDF Multi-wavelength catalogue was obtained via the Spitzer Large Area Survey with Hyper-Suprime-Cam \citep[SPLASH;][]{Mehta2018}. The million-row dataset contains corresponding photometric redshifts and over 10,000 spectroscopic redshifts, with relevant morphological information for each object. The CFHQSIR survey is a Y-band extension of the Canada-France-Hawaii Telescope Legacy Survey containing data from MegaCam and WIRCam \citep{Pipien2018}. This catalogue contains 690,000 rows with no redshift coverage but provides morphological information for faint ($Y \leq 23$\,mag) galaxies. Together, these catalogues summarise 1.5 million galaxies with extensive morphological coverage and some redshift information across the XMM-LSS Deep Drilling Field (Figure \ref{fig:0510}). 

Lestrade imports the DES sample data within XMM-LSS and constructs the master catalogue described above. For each transient, every potential host within a given radius (which we will tune) 
is identified with a corresponding $d_{\mathrm{DLR}}$ (or A-Value) calculated. These values are then sorted, with the best and second-best hosts identified as those with the lowest and second-lowest $d_{\mathrm{DLR}}$. Here, the pipeline checks the proximity of these two hosts -- if their shapes are found to have significant overlap (such that the two morphologies may describe the same or a sufficiently blended host), the second-best host is discarded, and the next-best host is selected to remove duplicate catalogue entries. A real and unique second-best host provides valuable context in determining the accuracy of our best choice. 

In cases such as our DES sample, where the true (DES-identified) host is known, the code then inspects its best choice against the coordinates of the `true' host. If the coordinates fall within the ellipse of our selected host (or within a given distance of the ellipse edge), we designate it a successful match (or at least consistent with the prior selection by DES). If the matches are inconsistent, we designate the attempt as unsuccessful. When the true host coordinates do not fall within any galaxy in our catalogue, we classify it as an `Incomplete' case and exclude it from the study. The user may then examine each result and choose to override any case as they see necessary. In the case of XMM-LSS, we identify 22 `inconsistent' matches for inspection. These were a mixture of cases where a) the `true' host coordinates lay just outside the best identified host (i.e. the co-ordinates are less than half an arcsecond outside the ellipse), and b) the `true' host is so far away from the transient (multiple arcminutes away) that we state the closer host identified by our code is a better match. 

\begin{figure}
\centering
 \includegraphics[width=1.0\columnwidth]{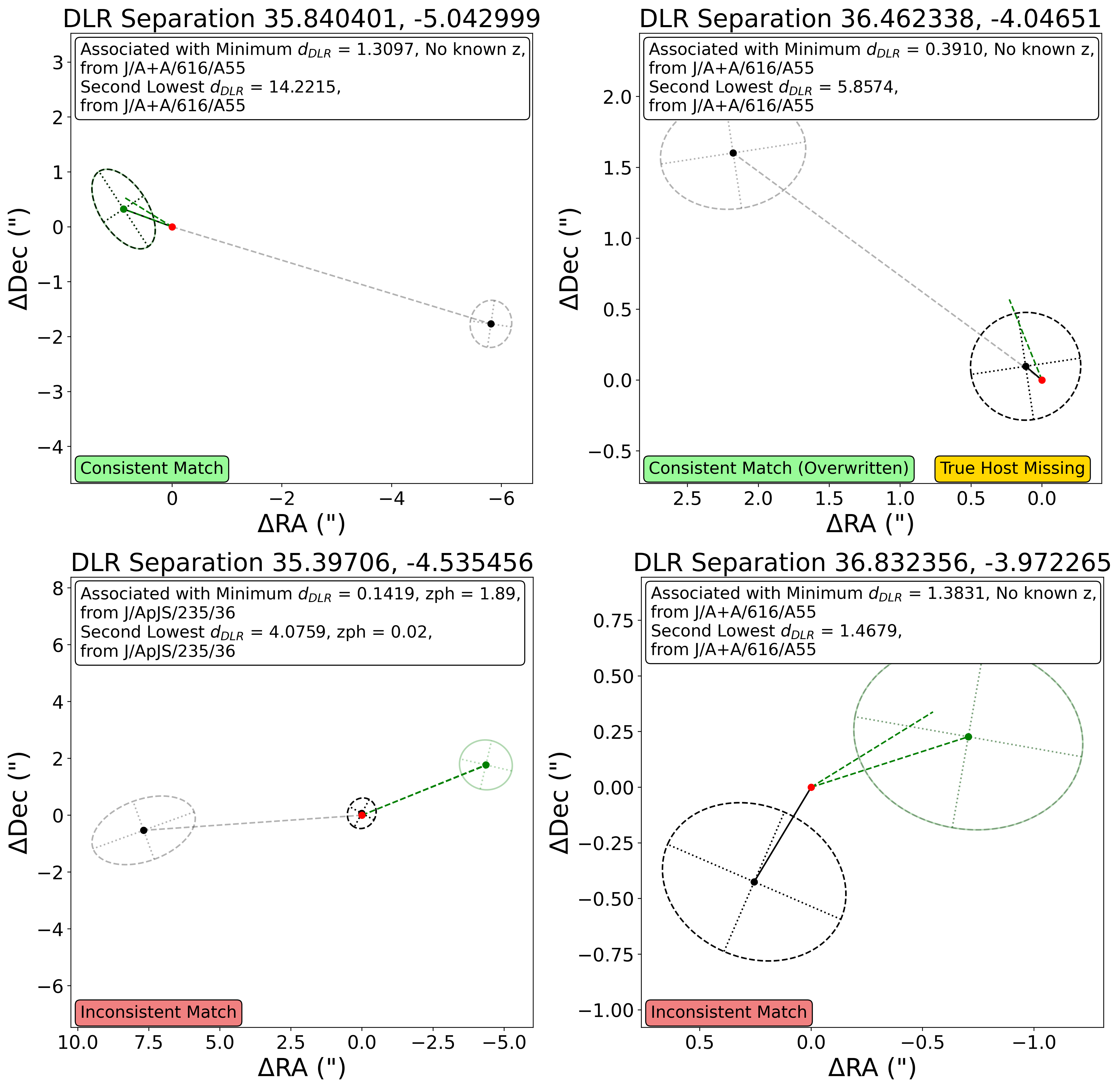}
 \caption{Lestrade visual aid plots for transient-host matching with the DES sample in XMM-LSS. The red marker denotes the transient location, with the DLR method's identified best host (black) and second-best host (grey). A green dashed line links to the true host coordinates in the DES sample, and the morphology if the host exists within the catalogue. Top left: A consistent match by the DLR method to the host identified by DES. Top right: A consistent match by the DLR method. In this case, the match has been manually labelled as correct following visual inspection. Bottom left: An inconsistent match by the DLR method where the true host was neither identified as the best nor the second-best candidate host. Bottom right: An inconsistent match by the DLR method where the true host was identified as the second-best host.}
 \label{fig:0511}
\end{figure}

Upon selecting the two best hosts for each transient and verifying the match, we output the results in both tabular form and as individual plots (see Figure \ref{fig:0511}). These plots provide a simplified visual of the relationship between the transient and the host, with additional contextual data such as redshift, redshift type, and the host source catalogue. Via inspection, we select certain cases for further annotation. Blended hosts, for example, are unsuitable for testing cross-matching methods; without follow-up observations, it is near-impossible to determine the true transient host, and so we cannot use these cases to compare the DLR and A-Value methods. We use the visual aids to identify them in our results and remove them from subsequent analysis.

\begin{figure}
\centering
 \includegraphics[width=1.0\columnwidth]{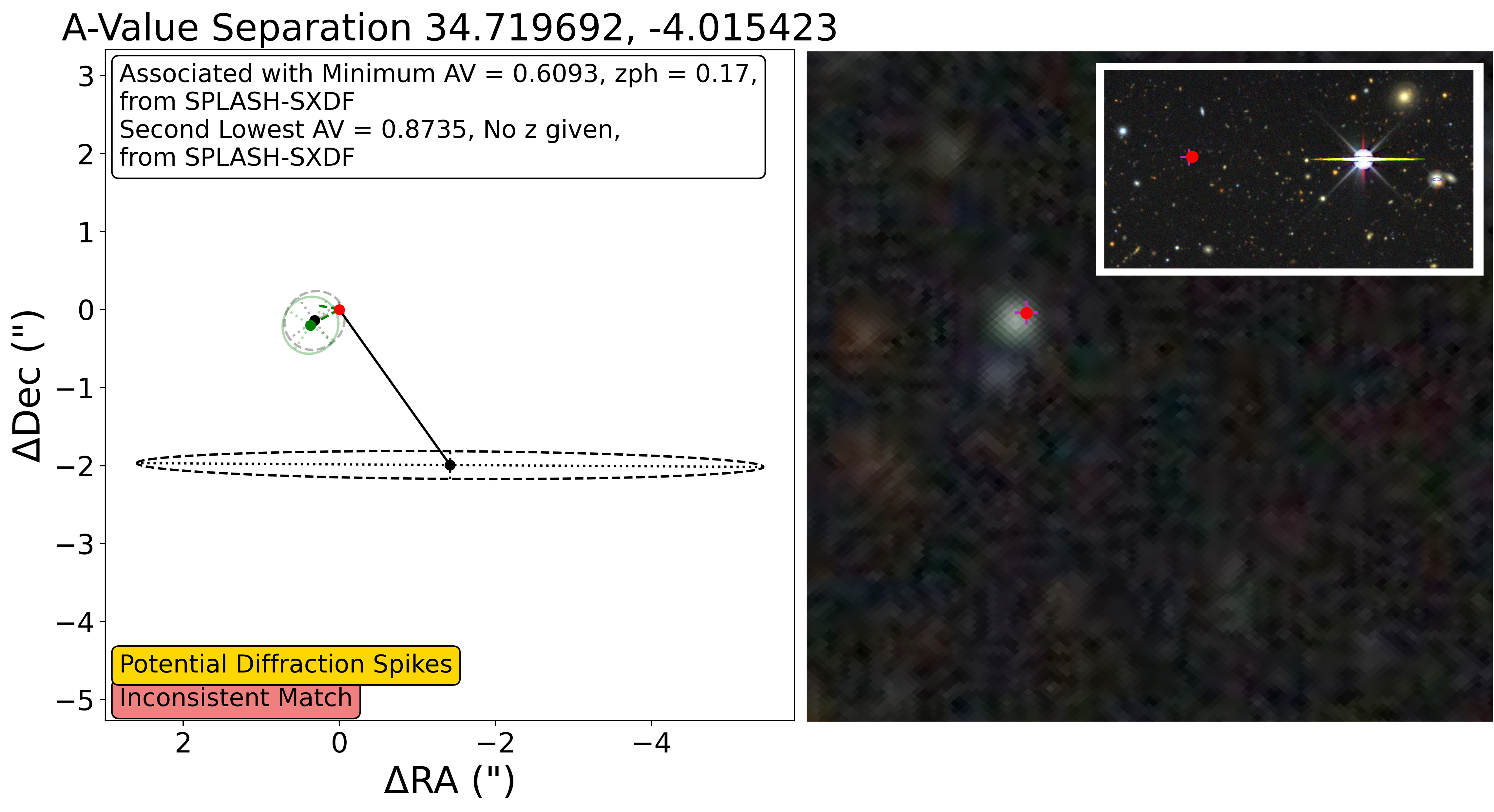}
 \caption{Diffraction spike source identified in an A-Value match. Note the large bright source along the axis of the diffraction spike (inset); a rotating variable (BD-04 376) with magnitude V = 10.05.}
 \label{fig:0513}
\end{figure}

\begin{figure}
\centering
 \includegraphics[width=1.0\columnwidth]{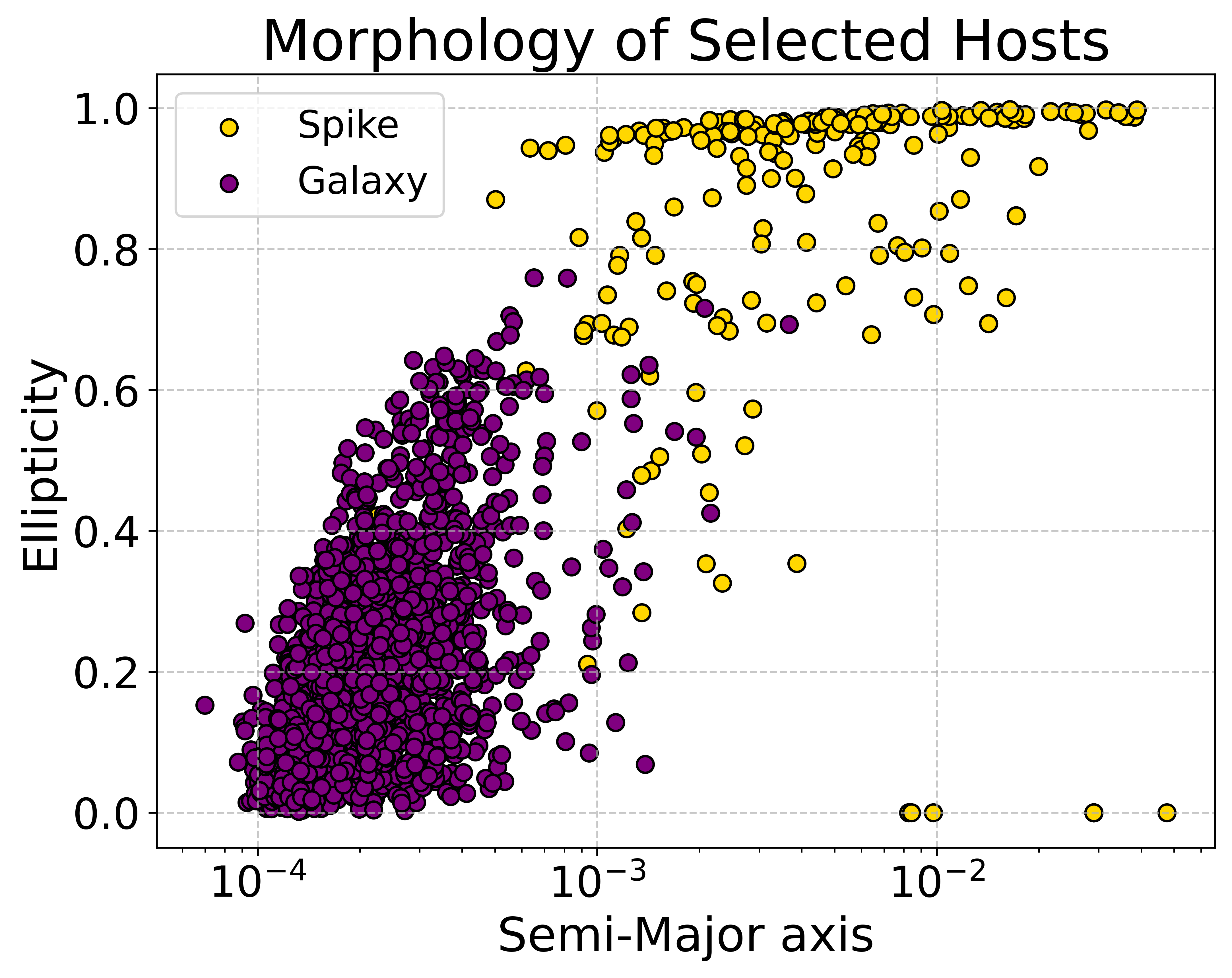}
 \caption{Diffraction spike sources identified in initial application of the DLR and A-Value methods.}
 \label{fig:0512}
\end{figure}

Another larger subgroup in our results has identified hosts in the catalogues with very large semi-major axes and high ellipticity. On examination of survey image data, these `hosts' have been identified as diffraction spikes -- see Figures \ref{fig:0512} and \ref{fig:0513}. These spikes need to be removed to make the best use of the catalogues. The large semi-major axes of these spikes provide very low  $d_{\mathrm{DLR}}$ values not directed towards the transient, `beating' closer, more realistic hosts. For the A-Value method, where ellipticity is ignored, this problem is amplified as a large, arcminutes-scale `galaxy' dominates over all galaxies in its region. We identify these again via the visual aids in tandem with survey image data. We perform a rudimentary culling of diffraction spikes from our XMM-LSS master catalogues (SPLASH-SXDF and CFHQSIR) by deleting rows with ellipticity greater than 0.8 and angular size greater than $10^{-3}$, and rerun the same pipeline. The eighteen real galaxies removed by this cut are large enough and bright enough (K $\sim12$) to be found in the larger catalogues already within the Sherlock database.

\subsection{Performance Comparison}
\begin{figure}
\centering
 \includegraphics[width=0.6\columnwidth]{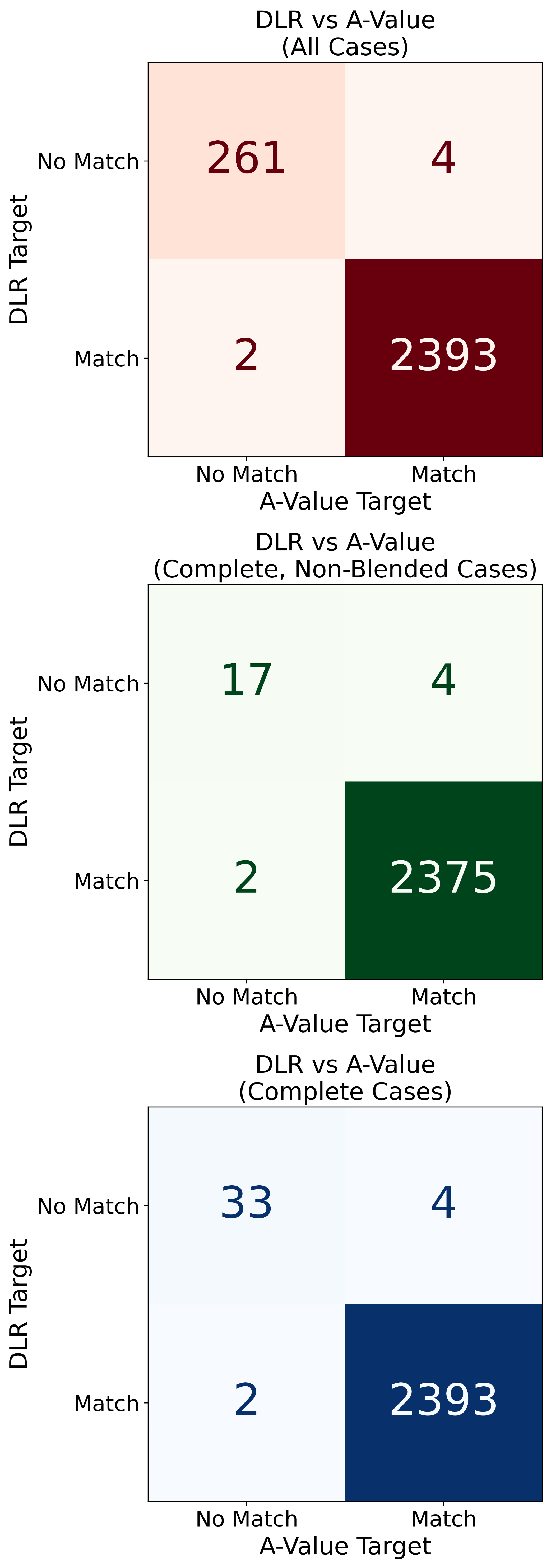}
 \caption{Confusion matrices comparing DLR and A-Value approaches for DES transients in XMM-LSS on catalogues where more extreme diffraction spikes have been culled.}
 \label{fig:0514a}
\end{figure}

Confusion matrices are produced for the DES XMM-LSS Sample as seen in Figure \ref{fig:0514a}. We provide the matrices for both the `cleaned' sample (with incomplete cases and blended hosts removed) and the `uncleaned' sample (with no cases removed). In an ideal case with a clean catalogue and exhaustive coverage, these matrices would converge to near or complete equivalence; we provide both to showcase the impact of `indeterminable cases' (or blended hosts and incomplete cases). Across all cases, we see both methods fail to identify approximately 10\% of hosts. The DLR method fails to identify 4 hosts where the A-Value method is successful, while the A-Value method fails to identify 2 hosts where the DLR method is successful. Looking at the filtered confusion matrix, we see 2 of these A-Value specific successes are in fact diffraction spikes spanning the area of the true host. For the unique successes of the A-Value method, one case involves the true host being a distant but extended host that would have benefited from being treated as `larger'. In the second case, the A-Values for the best and second-best candidates are 2.9057 and 2.9455, with DLR values of 3.4030 and 3.0322, respectively. For the unique DLR successes, both galaxies have a similar angular distance, and benefit from the attention to detail provided by the approach.

\begin{figure}
\centering
 \includegraphics[width=0.6\columnwidth]{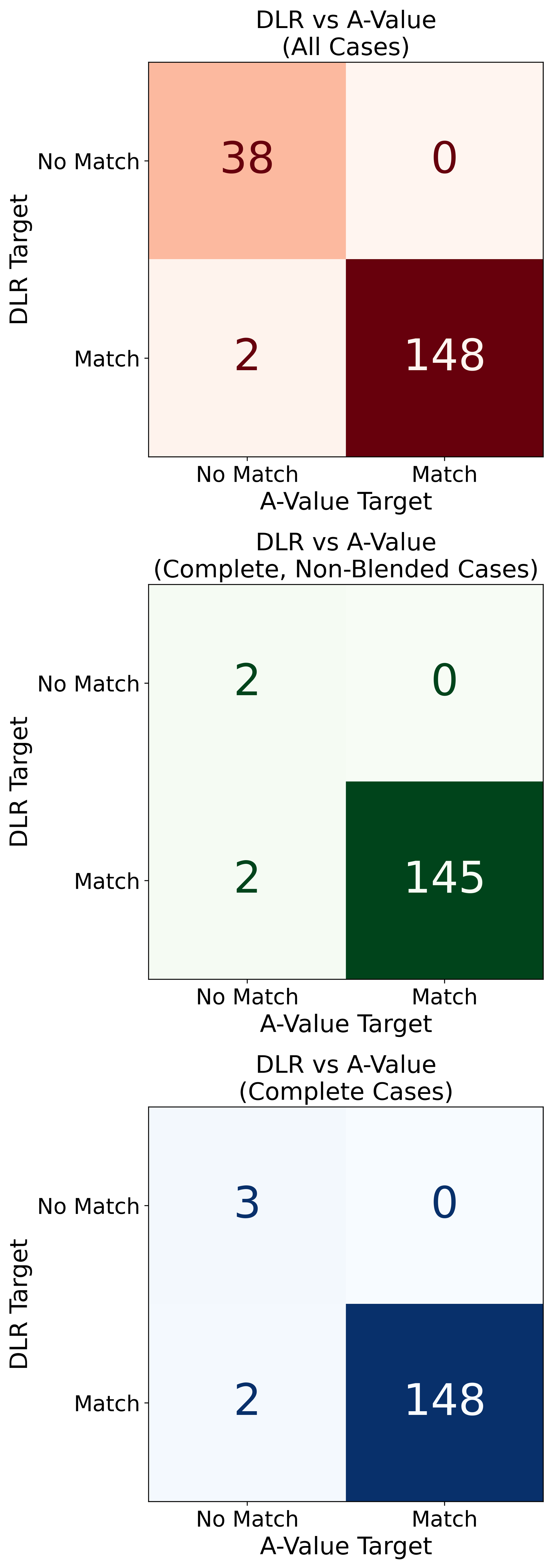}
 \caption{Confusion matrices comparing DLR and A-Value approaches for DES transients in ECDFS on catalogues where more extreme diffraction spikes have been culled.}
 \label{fig:0514b}
\end{figure}

We also have sufficient coverage in the ECDF-S field to compare the DLR and A-Value methods. This test is carried out on a smaller subsample of 188 DES transients. While we have a large number of morphological catalogues for this DDF, the majority of the coverage is limited to within half a degree of the central pointing. Subsequently, we only use the Chandra Deep Field South multi colour dataset \citep{Wolf2004}. The set contains the requisite morphological information (semimajor and semiminor axes and position angle) and photometric redshifts for each object. As before, we produce confusion matrices for the field using the pipeline outlined above (Figure \ref{fig:0514b}). The DLR method slightly outperforms the A-Value method with two unique successful matches by the former compared to 0 by the latter.

In the case of the DDFs, where the number of large hosts is limited, we can state that while the DLR method may examine morphologies to greater detail than the A-Value method, there is insufficient evidence to suggest a statistically significant improvement. What is apparent is that the DLR method, being more sensitive to the ellipticity and angle of a potential host, provides greater robustness against contaminants such as diffraction spikes than the A-Value approach. As with all morphology-focused approaches, both methods struggle with blended hosts; the visual aid plot will help users identify these cases, but the pipeline is limited in its ability to classify them automatically. The DLR approach may also be well-suited to scenarios where galaxy semi-major axes are similar and at similar angular separations from the transient. For XMM-LSS we find that 90\% of transients that could not be successfully matched did not contain the correct host in our catalogues, and 92\% for ECDF-S. As a greater number of surveys focus on the DDFs - LSST included - the number of incomplete cases should decrease, and may already be decreasing by implementing more catalogues or other host-matching approaches as outlined in Section 3.

\subsection{Cross-Matching Confidence Scores}
Given comparable performance from both the DLR and A-Value approaches, it is worth examining the success of each method in greater detail. We have manually inspected each case to highlight types of failures, such as diffraction spikes and `incomplete' cases. We have also overwritten successful cases which were initially automatically classed as incorrect. This annotated set of host matches and their outcomes - successful or unsuccessful - can be used to train a machine learning binary classification model that will assign a `confidence' score for each match made. The better the two classes are separated by each approach (DLR and A-Value) in the score distribution of our samples, the greater the confidence we can place in each method. Provided a well-performing model is trained for each model, we can reduce the time spent manually inspecting host matches by automatically classifying unlabelled host candidates as `successful' or `unsuccessful', either in real-time or offline.  

For our analysis, we continue to use the DES sample in the XMM-LSS DDF. We keep the overwritten cases within the sample as successes and do not remove incomplete cases; there is a benefit in all types of unsuccessful matches being included in the training set for the model, even if we try to avoid such cases in the pipeline. For the same reason, we retain blended hosts, though we expect their impact on our training to be minimal, as the group's volume remains small compared to other successful/unsuccessful cases.

\begin{figure}
\centering
 \includegraphics[width=1\columnwidth]{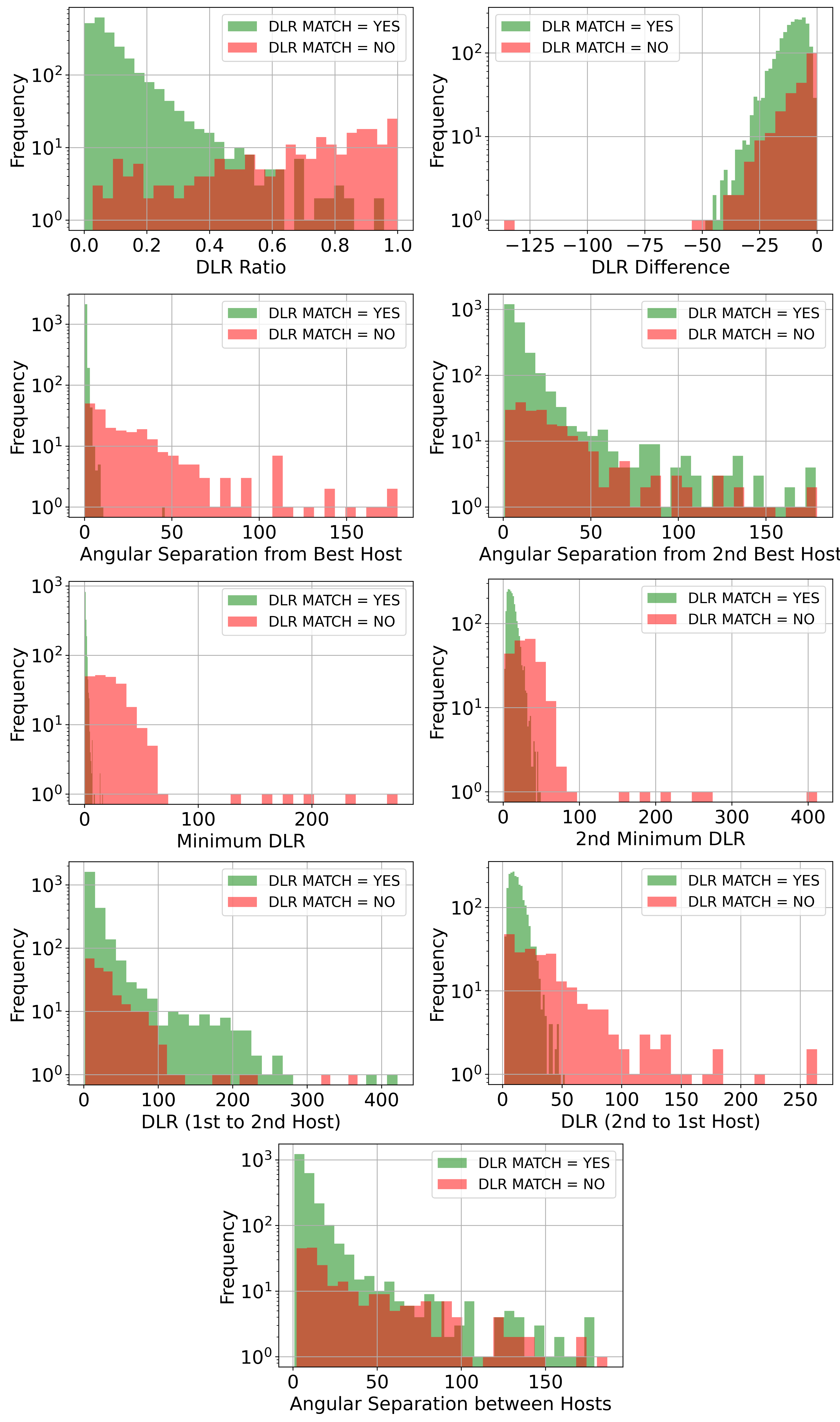}
 \caption{Feature distributions of successful/unsuccessful matches made by the DLR method.}
 \label{fig:0521}
\end{figure}

\begin{figure}
\centering
 \includegraphics[width=1\columnwidth]{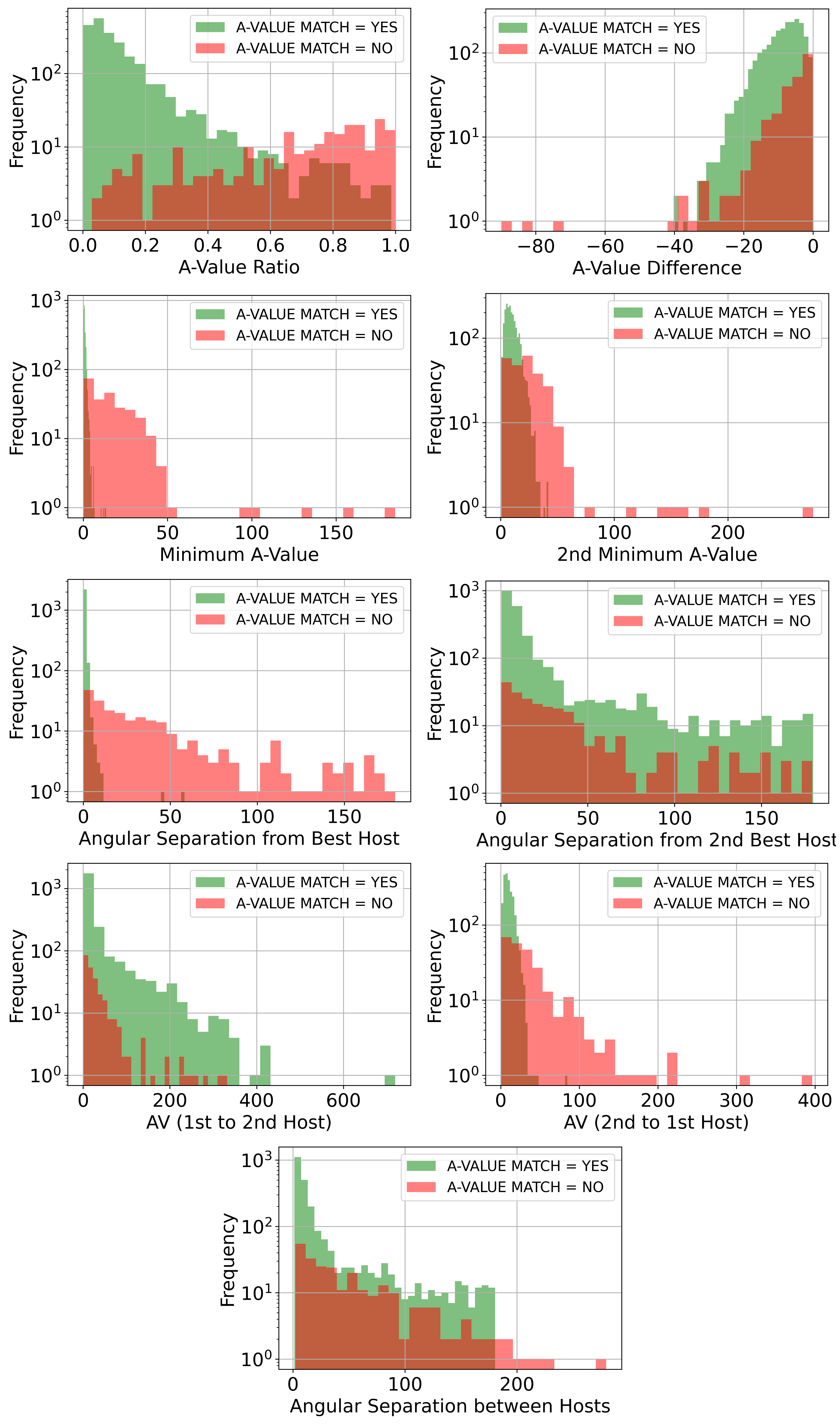}
 \caption{Feature distributions of successful/unsuccessful matches made by the A-Value method.}
 \label{fig:0522}
\end{figure}

We select a subset of features obtained during cross-matching for analysis, plotting histograms of their distributions by class (see Figure \ref{fig:0521}). For the DLR method, we identify several useful features: above a $d_{\mathrm{DLR}}$ of 20, all matches were unsuccessful, with most successful cases occurring at $d_{\mathrm{DLR}} < 10$. For the second minimum $d_{\mathrm{DLR}}$ we see all cases at $d_{\mathrm{DLR}2} > 50$ are unsuccessful. For the angular separation between the transient and the selected best host, we see no successful matches above 15 arcseconds. Finally, the $d_{\mathrm{DLR}}$ of the second host in the direction of the best host also provides a clear difference in the distributions between successful and unsuccessful matches, with no value of $d_{\mathrm{DLR}2\rightarrow1}$ above 50 corresponding to a success. We see similar cases in the A-Value approach for the minimum $A$, the angular separation between the transient and the best host and the A-Value of the second best host in relation to the centre of the first $A_{2\rightarrow1}$ (Figure \ref{fig:0522}). 

For our machine learning confidence model, we employ the XGBoost (extreme gradient boosting) model found in the scikit-learn Python package. An XGBoost model is a regularising gradient-boosting algorithm that sequentially builds decision trees to arrive at an optimised process. We divide our data into a training and testing set, reserving 90\% of our sample for the testing process, and train the model on the remaining 10\%, using our pre-selected features. The XGBoost approach successively trains decision trees to correct mistakes by the preceding tree over a fixed number of boosting rounds (in this case, 100) by minimising the logistic loss:

$$
\mathcal{L} = -\frac{1}{N} \sum_{i=1}^{N} \left[ y_i \log(\hat{y}_i) + (1 - y_i) \log(1 - \hat{y}_i) \right]
$$
Where $y_i$ is the true label (0 or 1), $\hat{y}_i$ is the predicted probability for class 1, and $N$ is the number of samples. The final model assigns a score to each given set of host-matching parameters as a prediction of confidence: 1 indicating a successful match and 0 indicating an incorrect match. We run the trained model on our testing set, plotting model score against frequency and labelling by true population label. 

Following a successful model training, feature importance metrics can be computed as the number of times each feature is used in any split across all trees within the model. Initially we utilise the method's best and second-best distance value ($d_{\mathrm{DLR}}$ or $A$), the difference ($d{\mathrm{DLR}}$ and $d{\mathrm{A}}$) and ratio ($r{\mathrm{DLR}}$ and $r{\mathrm{A}}$) between these values, and the angular separation between the transient and selected host (S) metric. Given the need for efficient real-time host association of transients in LSST, we want to reduce the size and complexity of our models, and therefore reduce redundant metrics. Ranking the inputs by feature importance for our `alpha' models, we see a clear emphasis on value in the angular separation and minimum distance value, with the difference between the two lowest distance values in third. For our final inputs, we select these values alone, which, in both the DLR and A-Value cases, appear to improve model performance. 

\begin{figure}
\centering
 \includegraphics[width=1\columnwidth]{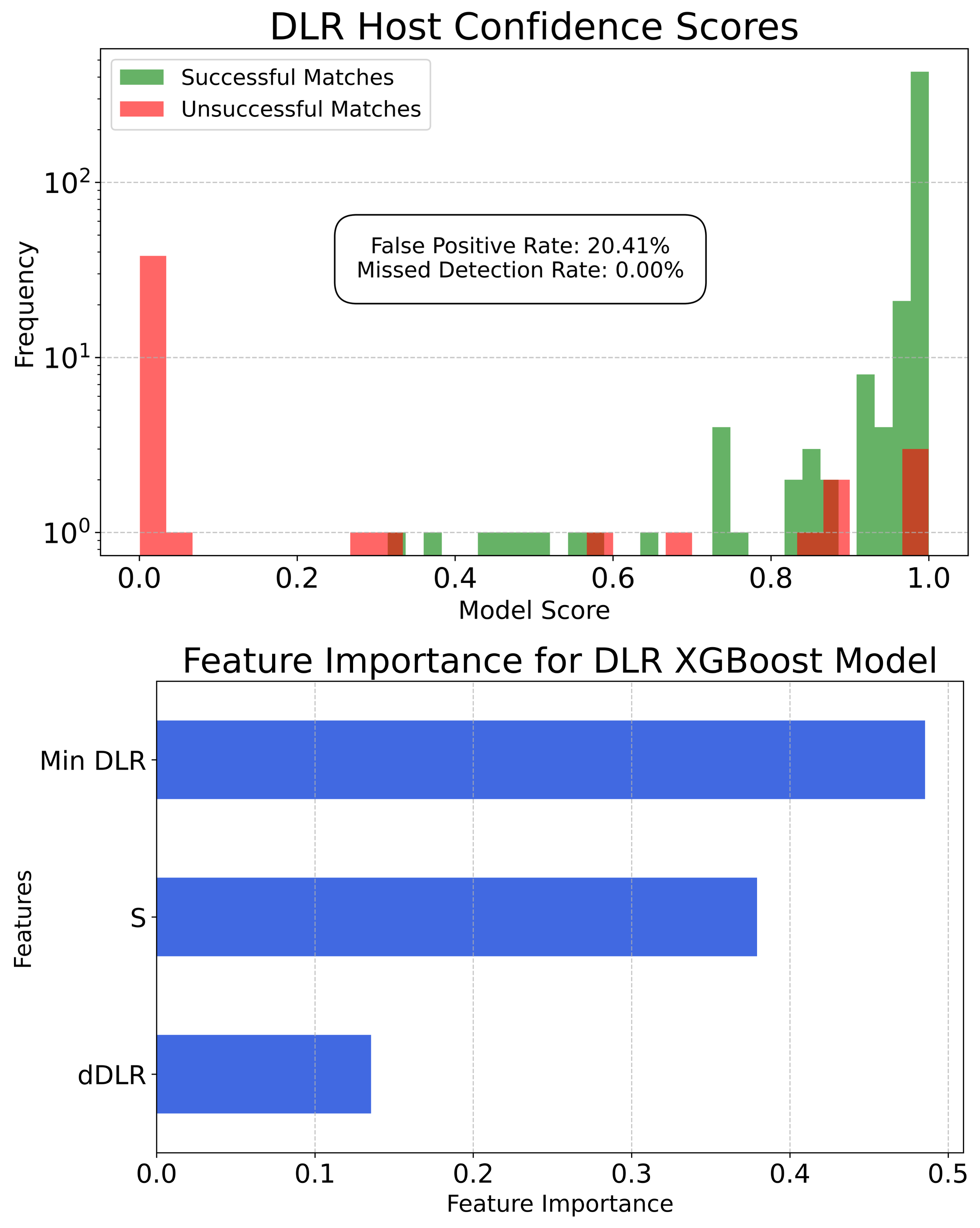}
 \caption{Score distribution of successful/unsuccessful matches by the DLR Confidence model.}
 \label{fig:0523}
\end{figure}

Examination of the DLR confidence model output yields promising results, with the majority of successful and unsuccessful cases assigned scores at either end of the score distribution, with only 18 out of 532 in the test set assigned scores between 0.2 and 0.8 (\ref{fig:0523}). A score threshold of 0.2, above which every case is flagged as successful and below which every case is flagged as unsuccessful, correctly classified each successful match. We can define our model performance by the false positive rate (FPR) and missed detection rate (MDR):

$$
FPR = \frac{FP}{FP + TN} \times 100\%
$$

$$
MDR = \frac{FN}{FN + TP} \times 100\%
$$

The former refers to the percentage of incorrect matches (False Positives, FP, and True Negatives, TN) that are incorrectly selected as successful, while the latter refers to the percentage of correct matches (False Negatives, FN, and True Positives, TP) that are erroneously selected as unsuccessful. For a score threshold of 0.2, we have a missed detection rate of 0\%, with only 10 false positives in total corresponding to a 20.41\% false positive rate. 

We manually examine these false positives one by one. The lowest-scoring false positive has no information regarding its true host in our catalogued information, and remains reasonably far from the assigned best host (some 25 arcseconds away), explaining the low confidence. The minimum $d_{\mathrm{DLR}}$ of 6.58 likely prevents the match from being scored lower, with a $d_{\mathrm{DLR}}$ difference of 23.87 indicating no other nearby hosts exist in the catalogue data aside from the best selected candidate, thus boosting the score. The next-lowest-scoring false positive is due to a small galaxy lying directly between the transient and the true host. After this, the next false positive is a blended host pair. Assuming the two blended galaxies will always be selected as the best and second best galaxies, they will always maintain a low angular separation, minimum $d_{\mathrm{DLR}}$ and $d_{\mathrm{DLR}}$ difference, leading to a higher score. In such a case, a high score is acceptable as a ``semi-successful" match. The remaining cases are due solely to failures of the $d_{\mathrm{DLR}}$ method itself - the transient is closer to another potential host, or within a different potential host, and so has been selected owing to the lower $d_{\mathrm{DLR}}.$ These cases are unavoidable for a solely morphology-based approach and can only be corrected by including other methods, such as a redshift-based approach or taking host magnitude into account.

\begin{figure}
\centering
 \includegraphics[width=1\columnwidth]{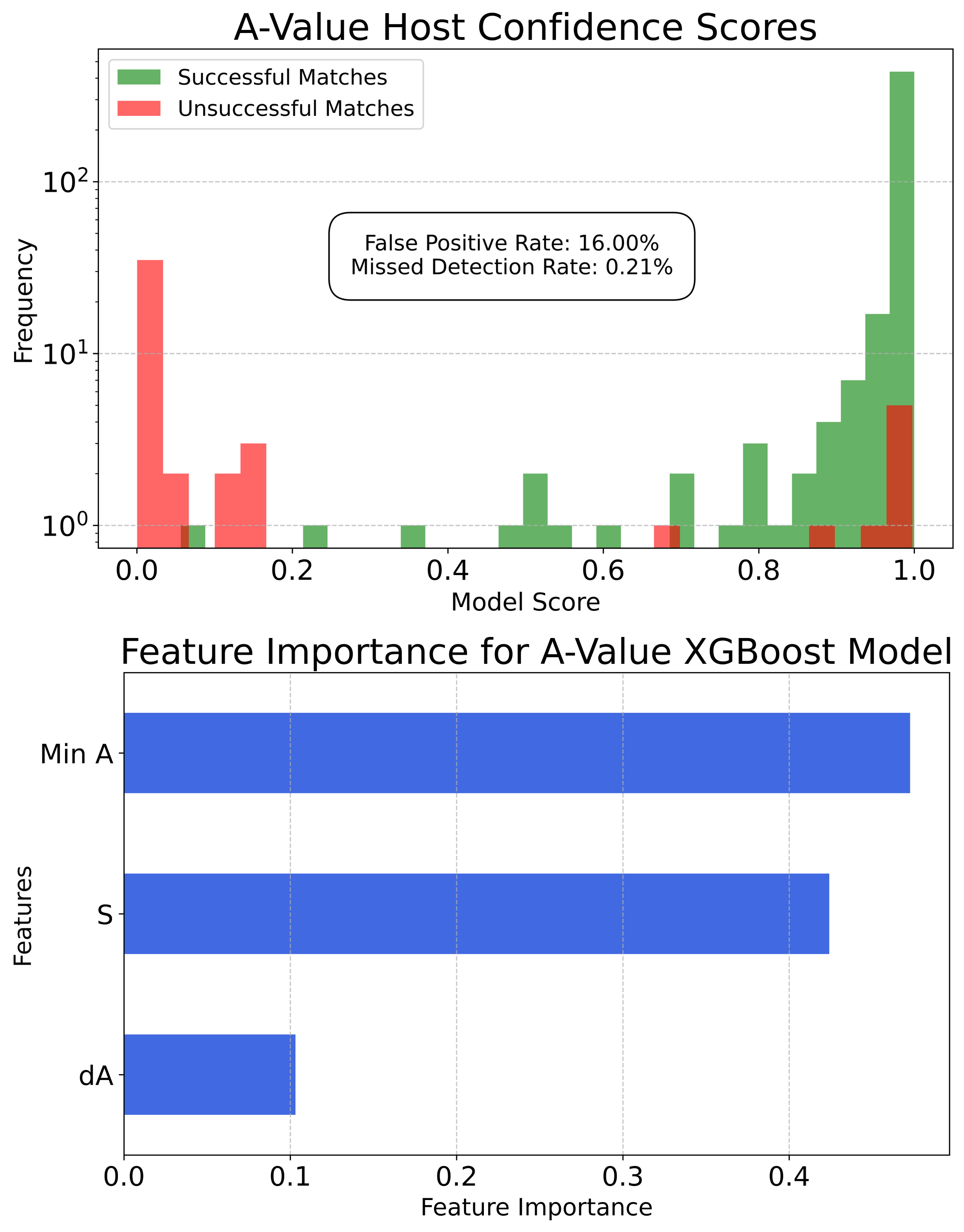}
 \caption{Score distribution of successful/unsuccessful matches by the A-Value Confidence model.}
 \label{fig:0524}
\end{figure}

The A-Value metrics are less useful in separating successful and unsuccessful cases (Figure \ref{fig:0524}). For the same score threshold of 0.2, we now have 8 false positives corresponding to an FPR of 16.00\%. We also have 1 missed detection corresponding to an MDR of 0.21\%.

Of the ten false positives identified in the DLR confidence model, seven appear in the false positive list for the A-Value model, excluding the case for which no true host exists, and two cases where the hosts have similar angular separations. Notably, the blended host case is scored higher by the A-Value model, whereas the DLR model appears to exert more caution. For the A-Value false negative, the true host lies approximately 4 arcseconds away and the second-best host 6 arcseconds away. Both galaxies have small semi-major ($\sim2$ arcseconds) and semi-minor ($\sim1$ arcsecond) axes.

While the DLR model is more successful in discerning successful matches from unsuccessful matches, it should be noted that this is not an indicator of the performance by either approach: it is simply an indicator of which method will be easier to manually inspect. As the DLR model more reliably scores successful matches highly and rarely provides a low score to a successful match, we can spend less time manually inspecting each cross-match. Should the morphology-based approach be combined with other approaches (such as the redshift-based approach), the model score can be used to determine how well supported a potential host is by its morphological information, used in tandem with scores or metrics from other data.

\subsection{Cross-Matching in the Wider Sky}
Galaxies in the DDF catalogues are mostly at greater distances than those in wide-field catalogues, and therefore have a smaller angular size. This makes it difficult to resolve their shape, as they may appear comparable to or smaller than the point-spread function of the instrument used to image them. For nearby galaxies with a larger angular extent and more detailed morphology, the difference between the DLR and A-Value approaches might be even more apparent.

\begin{figure}
\centering
 \includegraphics[width=0.6\columnwidth]{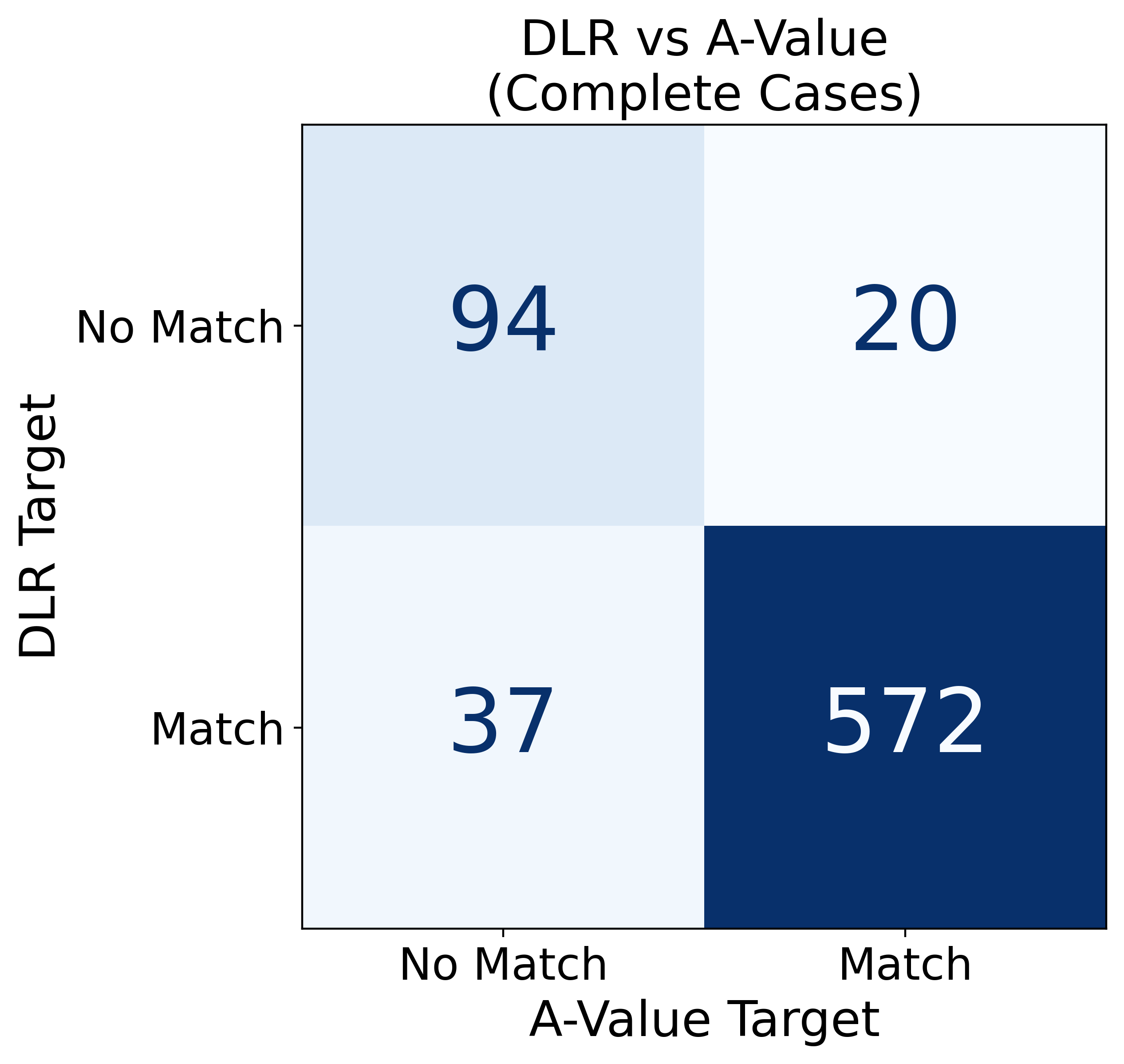}
 \caption{Confusion matrices comparing DLR and A-Value approaches for ATLAS transients on the Sloan Digital Sky Survey.}
 \label{fig:0531}
\end{figure}

To test this, we use the ATLAS100 sample of transients discovered by ATLAS (Srivastav et al., in prep) \citep{Smith_2020} with known host galaxies, and match them against the Sloan Digital Sky Survey (SDSS) Data Release 16 catalogued galaxies \citep{Ahumada_2020}. We find that of 723 transients in the sample for which the true host is in SDSS, 572 are successfully matched by both methods. 37 are matched by only by the DLR, and 20 only by the A-Value method (Figure \ref{fig:0531}). Inspection of each case shows a number of SDSS contaminants, including knots in the spiral arms of large galaxies being identified as individual galaxies. Although at times both methods select these knots rather than the true galaxy, we see the DLR method as being more robust against these contaminants. While an attempted removal of some of these galaxy components yielded a more similar performance between the two methods, the perfect implementation of these morphological approaches for wider-sky cross-matching is beyond the scope of this study. When galaxies are better resolved, a combination of methods is required to identify the correct host galaxy. Among the morphological techniques discussed, the DLR method remains the better-performing approach in both DDFs and wide-sky applications, even with contaminated data.

\section{Implementation}
The initial implementation priority is to install the A+ and A-ranked catalogues with more than 10,000 rows into the Sherlock database. Owing to the small size of these catalogues in relation to the currently stored data ($\sim0.1$\%) the true challenge to implementation is the human labour required to translate the original catalogue schema for use by Sherlock. For COSMOS the selected catalogues - the COSMOS/Ultravista and the FourStar Galaxy Evolution survey - contain a combined 296,079 rows. For XMM-LSS, FourStar and VIPERS contain 351,771 rows. The MUSYC Photometric Redshift Catalogue and FourStar data in ECDFS provide 136,431 rows for this field. As no A or A+ ranked catalogues have been found for ELAIS-S1, we are left with approximately 784,000 rows across four catalogues to implement for three of the DDFs. In future, we will expand this implementation to lower-tier catalogues.

It is ultimately difficult to select an optimal method between the DLR and A-Value approaches. Both maintain a level performance in the DDFs, with a low number of unique successes and failures (and a low number of mutual failures overall). From a solely morphology-based perspective, either approach is optimal.

When implementing real-time transient-host matching, the trade-off between computational expense and performance must be considered. The DLR method involves more pipeline operations and requires more catalogued information, such as the galaxy position angle and semi-minor axis. While not a significant constraint for offline matching, this will compound as more transients are processed in the Rubin pipeline. Running the Sherlock software 21 times on a set of $\sim24,000$ transients, we found that the classification rate with the DLR method implemented was 418.5 transients/sec, with a DLR calculation overhead of 18 microseconds/transient, increasing the Sherlock runtime by only 0.8\%. Given the minimal contribution of the DLR calculation to the total runtime, the method has now been integrated into Sherlock. 

We must also consider how each method handles `contaminated' catalogues. The DLR method is more robust against contaminants and less likely to misidentify diffraction spikes as real objects, whereas the A-Value approach is more susceptible to artefacts. These contaminants should be removed prior to catalogue implementation, though some may survive culling and only be identified once selected as a potential host. If contaminants are expected in some or all of the catalogued data, the DLR method will be more successful at identifying true hosts.

\section{Conclusion}
In this paper, we have explored a method for analysing and utilising existing high-quality galaxy catalogues for transient-host matching in LSST's DDFs. We identify catalogues from various surveys across four DDFs, with a large number of photometric and spectroscopic redshifts, morphology measurements and object type flags. These analyses and the requisite code have been uploaded to a repository wiki as a resource for transient scientists to identify useful catalogues. 

We compare the Directional Light Radius method with the A-Value method of transient host-matching and test their performance on a pre-assessed sample of DES transients in the XMM-LSS and ECDFS fields. In carrying out our own host association, we identify a similar performance between each approach, but highlight that the DLR approach is more robust against large-scale contaminants such as diffraction spikes. We discuss other benefits of each method, such as the A-Value method's advantage over poorly-fitted morphologies and the DLR method's advantage for evenly-separated galaxies. 

We examine machine learning-based avenues for assessing host selection confidence, again focusing on the DLR and A-Value methods. We find that the DLR confidence model correctly identifies successful matches with a lower missed-detection rate than the A-Value model. While this should not be used to promote one approach's inherent superiority, it highlights a way of identifying successful matches for unlabelled data.
We note the difficulty caused by catalogue contamination from artefacts (such as diffraction spikes) when assessing host-matching methods, and the problem in identifying the true host between a blended pair.

Finally, we look to integrate the A+ and A-ranked catalogues into Sherlock for host cross-matching in the DDFs by Lasair, which contain 784,000 sources in total across COSMOS, XMM-LSS and ECDFS, with additional catalogues to be added later. Implementing the DLR method within Sherlock contributes very little to the total runtime (0.8\%) at a rate of $18\mu$s per transient. The DLR method is now implemented within Sherlock.

An avenue for future research would be to formally assess the performance of a redshift-based host-matching approach compared to our morphology-based approach, or to a simple angular-separation-based approach. Combining these approaches and incorporating a host confidence model, as we have begun exploring above, may be a useful exercise. 

We also note a lack of publicly available and accurate transient-host samples for testing transient science in the LSST DDFs. The construction of such a sample would prove useful in late-stage preparations for real-time transient science with Rubin.

\begin{acknowledgments}
JW acknowledges a studentship funded by the Leverhulme Trust through the Leverhulme Interdisciplinary Network on Algorithmic Solutions (LINAS) at Queen's University Belfast. MJJ  acknowledges the support of a UKRI Frontiers Research Grant [EP/X026639/1], which was selected by the European Research Council. MJJ, IHW, SJS and KWS acknowledge support from the Oxford Hintze Centre for Astrophysical Surveys which is funded through generous support from the Hintze Family Charitable Foundation. MN is supported by the European Research Council (ERC) under the European Union’s Horizon 2020 research and innovation programme (grant agreement No.~948381).
SJS, DRY and KS acknowledge funding from STFC Grants ST/X001253/1, ST/Y001605/1, and a Royal Society Research Professorship. 

This work has made use of CosmoHub, developed by PIC (maintained by IFAE and CIEMAT) in collaboration with ICE-CSIC. It received funding from the Spanish government (grant EQC2021-007479-P funded by MCIN/AEI/10.13039/501100011033), the EU NextGeneration/PRTR (PRTR-C17.I1), and the Generalitat de Catalunya.
\end{acknowledgments}

\bibliography{References}{}
\bibliographystyle{aasjournalv7}

\appendix

\section{Additional Figures}
\begin{center}
\rotatebox{90}{%
  \resizebox{1.1\textheight}{!}{%
    \begin{tabular}{|l|l|r|r|r|r|r|r|r|l|l|l|}
    \hline
    \textbf{ID} & \textbf{Catalogue} & \textbf{No. Objects} & \textbf{zspec no.} & \textbf{zspec 90} & \textbf{zspec 50} & \textbf{zphot no.} & \textbf{zphot 90} & \textbf{zphot 50} & \textbf{Morphology} & \textbf{Type\_Flag} & \textbf{Rank} \\
    \hline
    \citep{Griffith2012} & The ACS-GC catalog & 304688 & 10013 & 1.42 & 0.61 & 251592 & 1.99 & 0.83 & Yes & Yes & A+ \\
    \citep{DESI2025} & DESI DR1 & 204977 & 199523 & 1.41 & 0.72 & & & & & Yes & A \\
    \citep{Muzzin2013} & COSMOS/UltraVISTA Ks-selected catalogs & 262615 & 5537 & 0.84 & 0.5 & 0 & & & & Yes & A \\
    \citep{Straatman2016} & FourStar galaxy evolution survey & 33464 & 465 & 3.07 & 0.73 & 0 & & & Yes & & A \\
    \citep{Lilly2007} & zCOSMOS-bright catalog, DR3 & 20689 & 19148 & 0.93 & 0.53 & & & & & & A- \\
    \citep{Hasinger2018} & DEIMOS 10K spectroscopic survey in COSMOS field & 10873 & 8786 & 3.1 & 0.89 & & & & & & A- \\
    \citep{Weaver2022} & The COSMOS2020 Farmer catalog & 2685207 & & & & 1439880 & 3.35 & 1.18 & Yes & Yes & B+ \\
    \citep{shuntov2025} & The COSMOS2025 Catalog & 784016 & & & & 681218 & 3.45 & 1.14 & Yes &  & B \\
    \citep{Darvish2017} & Cosmic web of galaxies in the COSMOS field & 45421 & & & & 45421 & 1.13 & 0.84 & & Yes & B \\
    \citep{Molino2016} & ALHAMBRA Survey & 38427 & & & & 38427 & 1.75 & 0.9 & Yes & & B \\
    \citep{Hatfield2022} & Photometric Redshifts in COSMOS & 995049 & & & & 995049 & 3.98 & 2.16 & & & B- \\
    \citep{Capak2007} & COSMOS Multi-Wavelength Photometry Catalog & 438226 & & & & 438226 & 1.98 & 0.74 & & & B- \\
    \citep{Razim2021} & COSMOS2015 dataset machine learning photo-z & 214398 & & & & 214398 & 1.09 & 0.75 & & & B- \\
    \citep{Nayyeri2017} & Multi-wavelength data in CANDELS COSMOS field & 38671 & 0 & 0 & & & & & Yes & & C \\
    \citep{Hale2025} & MIGHTEE DR1 & 20886 &  &  & & & & & Yes & & C \\
    \citep{McCracken2012} & The fourth UltraVISTA data release & 451587 & & & & & & & & & C- \\
    \citep{Hermes2017} & Herschel Multi-tiered Extragalactic Survey & 124343 & & & & & & & & & C- \\
    \citep{Chang2021} & Machine learning predicted AGNs in HSC-Wide region & 112609 & & & & & & & & & C- \\
    \citep{Cabayol2019} & Star-galaxy multi narrow-band classification & 38427 & & & & & & & & & C- \\
    \citep{Smolvcic2017} & VLA-COSMOS 3 GHz Large Project & 10830 & & & & & & & & & C- \\
    \hline
    \end{tabular}%
  }%
}
\end{center}

\captionsetup{hypcap=false}
\captionof{table}{Deep Drilling Field catalogues in COSMOS.}

\vspace*{3cm} 

\begin{center}
\rotatebox{90}{%
  \resizebox{1.2\textheight}{!}{%
    \begin{tabular}{|l|l|r|r|r|r|r|r|r|l|l|l|}
    \hline
    \textbf{ID} & \textbf{Catalogue} & \textbf{No. Objects} & \textbf{zspec no.} & \textbf{zspec 90} & \textbf{zspec 50} & \textbf{zphot no.} & \textbf{zphot 90} & \textbf{zphot 50} & \textbf{Morphology} & \textbf{Type\_Flag} & \textbf{Rank} \\
    \hline
    \citep{Chen2018} & XMM-LSS SERVS. New XMM-Newton point-source cat. & 396287 & 42985 & 0.99 & 0.63 & 390900 & 1.65 & 0.85 & & Yes & A \\
    \citep{Garilli2014} & VIMOS Public Extragalactic Survey (VIPERS) DR1 & 316717 & 13525 & 0.98 & 0.71 & & & & & Yes & A \\
    \citep{DESI2025} & DESI DR1 & 41119 & 39353 & 1.49 & 0.80 & & & & & Yes & A \\
    \citep{Cabayol2019} & The PAU Survey & 218280 & 29124 & 0.92 & 0.62 & 218280 & 1.01 & 0.58 & & Yes & A \\
    \citep{Straatman2016} & FourStar galaxy evolution survey & 35054 & 152 & 1.49 & 0.75 & 0 & & & Yes & & A \\
    \citep{Rowan-Robison2013} & Revised SWIRE photometric redshifts & 311643 & 3628 & 1.15 & 0.63 & & & & & & A- \\
    \citep{Mehta2018} & SPLASH-SXDF multi-wavelength photometric catalog & 1088181 & & & & 1084522 & 2.97 & 1.02 & Yes & Yes & B+ \\
    \citep{Moutard2016} & VIPERS Multi-Lambda Survey & 740712 & & & & 740160 & 1.65 & 0.75 & & Yes & B \\
    \citep{Hatfield2022} & Photometric Redshifts in XMM-LSS & 1674689 & & & & 1674689 & 4.07 & 1.98 & & & B- \\
    \citep{Lacy2021} & Spitzer Survey of Deep Drilling Fields & 694991 & & & & 692977 & 1.03 & 2.33 & & & B- \\
    \citep{McCracken2008} & VIRMOS deep imaging survey. VVDS-F02 catalog & 2028704 & & & & & & & Yes & Yes & C+ \\
    \citep{Jarvis2013} & VISTA Deep Extragalactic Observations Survey (VIDEO) & 1279857 & & & & & & & Yes & Yes & C+ \\
    \citep{Pipien2018} & CFHQSIR survey & 691419 & & & & & & & Yes & Yes & C+ \\
    \citep{Hale2025} & MIGHTEE DR1 & 53393 &  &  & & & & & Yes & & C \\
    \citep{Hermes2017} & Herschel Multi-tiered Extragalactic Survey & 258612 & & & & & & & & & C- \\
    \hline
    \end{tabular}%
  }%
}
\end{center}

\captionsetup{hypcap=false}
\captionof{table}{Deep Drilling Field catalogues in XMM-LSS.}
\vspace*{3cm} 

\begin{center}
\rotatebox{90}{%
  \resizebox{1.2\textheight}{!}{%
    \begin{tabular}{|l|l|r|r|r|r|r|r|r|l|l|l|}
    \hline
    \textbf{ID} & \textbf{Catalogue} & \textbf{No. Objects} & \textbf{zspec no.} & \textbf{zspec 90} & \textbf{zspec 50} & \textbf{zphot no.} & \textbf{zphot 90} & \textbf{zphot 50} & \textbf{Morphology} & \textbf{Type\_Flag} & \textbf{Rank} \\
    \hline
    \citep{Cardamone2010} & MUYSC Photometric Redshifts & 84402 & 3988 & 2.7 & 1.04 & & & & Yes & Yes & A+ \\
    \citep{Griffith2012} & The ACS-GC catalog & 70446 & 6955 & 2.02 & 0.69 & 44239 & 1.27 & 0.93 & Yes & Yes & A+ \\
    \citep{Straatman2016} & FourStar galaxy evolution survey & 52029 & 1226 & 2.67 & 0.97 & 0 & & & Yes & & A \\
    \citep{Grazian2006} & GOODS-MUSIC sample multicolour catalog & 18296 & 1275 & 1.34 & 0.73 & 14646 & 3.14 & 1.19 & & Yes & A \\
    \citep{Taylor2009} & MUSYC Survey & 16910 & 2914 & 1.23 & 0.68 & 13547 & 135.94 & 24.96 & & Yes & A \\
    \citep{Rowan-Robison2013} & Revised SWIRE photometric redshifts & 149766 & 728 & 1.1 & 0.52 & & & & & & A- \\
    \citep{Rafferty2011} & SMBH Growth in Starbust Galaxies with Chandra & 100318 & 2587 & 2.31 & 0.74 & 100318 & 3.44 & 1.62 & & & A- \\
    \citep{Lacy2021} & Spitzer Survey of Deep Drilling Fields & 799606 & & & & 787398 & 1.10 & 2.41 & & & B- \\
    \citep{Jarvis2013} & VISTA Deep Extragalactic Observations Survey (VIDEO) & 1138485 & & & & & & & Yes & Yes & C+ \\
    \citep{Wolf2004} & Chandra Deep Field South multi-colour data & 63501 & & & & & & & Yes & Yes & C+ \\
    \citep{Guo2013} & GOODS-S CANDELS multiwavelength catalog & 34930 & & & & & & & & Yes & C \\
    \citep{Hale2025} & MIGHTEE DR1 & 17866 &  &  & & & & & Yes & & C \\
    \citep{Hermes2017} & Herschel Multi-tiered Extragalactic Survey & 254266 & & & & & & & & & C- \\
    \hline
    \end{tabular}%
  }%
}
\end{center}

\captionsetup{hypcap=false}
\captionof{table}{Deep Drilling Field catalogues in ECDFS.}

\vspace*{3cm} 

\begin{center}
\rotatebox{90}{%
  \resizebox{1.2\textheight}{!}{%
    \begin{tabular}{|l|l|r|r|r|r|r|r|r|l|l|l|}
    \hline
    \textbf{ID} & \textbf{Catalogue} & \textbf{No. Objects} & \textbf{zspec no.} & \textbf{zspec 90} & \textbf{zspec 50} & \textbf{zphot no.} & \textbf{zphot 90} & \textbf{zphot 50} & \textbf{Morphology} & \textbf{Type\_Flag} & \textbf{Rank} \\
    \hline
    \citep{Rowan-Robison2013} & Revised SWIRE photometric redshifts & 145587 & 145587 & -99 & -99 & & & & & & A- \\
    \citep{Jarvis2013} & VISTA Deep Extragalactic Observations Survey (VIDEO) & 842337 & & & & & & & Yes & Yes & A- \\
    \citep{Hermes2017} & Herschel Multi-tiered Extragalactic Survey & 86779 & & & & & & & & & C- \\
    \hline
    \end{tabular}%
  }%
}
\end{center}

\captionsetup{hypcap=false}
\captionof{table}{Deep Drilling Field catalogues in ELAIS-S1}



\end{document}